\documentclass[utf8]{frontiersSCNS} 

\usepackage{url,hyperref,lineno,microtype,subcaption}
\usepackage[onehalfspacing]{setspace}
\usepackage{epsfig}
\usepackage{graphicx}

\usepackage{amsmath,amssymb,amsfonts}
\usepackage{algorithmic}
\usepackage{rotating}
\usepackage{textcomp}
\usepackage{xcolor}
\usepackage{multirow}

\def\keyFont{\fontsize{8}{11}\helveticabold }
\def\firstAuthorLast{Feng {et~al.}} 
\def\Authors{Huanghao Fengr\,$^{1}$, Mohammad H. Mahoor\,$^{2,*}$ and Francesca Dino\,$^{2}$}
\def\Address{$^{1}$Changshu Institute of Technology, Suzhou Changshu, China\\
	$^{2}$Computer Vision and Social Robotics Labarotory, Department of Electrical and Computer Engineering, University of Denver}
\def\corrAuthor{2155 E Wesley Ave, Denver, CO 80210, USA}

\def\corrEmail{fenghuanghao@cslg.edu.cn, mmahoor@du.edu, francesca.dino@cuanschutz.edu}

\begin{document}
	\onecolumn
	\firstpage{1}
	
	
	\title[Running Title]{A Music-Therapy Robotic Platform for \\Children with Autism: A Pilot Study\\} 
	
	\author[\firstAuthorLast ]{\Authors} 
	\address{} 
	\correspondance{} 
	
	\extraAuth{}
	
	\maketitle
	
	\begin{abstract}
	Children with Autism Spectrum Disorder (ASD) experience deficits in verbal and nonverbal communication skills including motor control, turn-taking, and emotion recognition. Innovative technology, such as socially assistive robots, has shown to be a viable method for Autism therapy. This paper presents a novel robot-based music-therapy platform for modeling and improving the social responses and behaviors of children with ASD. Our autonomous social interactive system consists of three modules. Module 1 provides an autonomous initiative positioning system for the robot, NAO, to properly localize and play the instrument (Xylophone) using the robot's arms. Module 2 allows NAO to play customized songs composed by individuals. Module 3 provides a real-life music therapy experience to the users. We adopted Short-time Fourier Transform and Levenshtein distance to fulfill the design requirements: a) "music detection" and b) "smart scoring and feedback", which allows NAO to understand music and provide additional practice and oral feedback to the users as applicable. We designed and implemented six Human-Robot-Interaction (HRI) sessions including four intervention sessions. Nine children with ASD and seven Typically Developing participated in a total of fifty HRI experimental sessions. Using our platform, we collected and analyzed data on social behavioral changes and emotion recognition using Electrodermal Activity (EDA) signals. The results of our experiments demonstrate most of the participants were able to complete motor control tasks with~70\% accuracy. Six out of the 9 ASD participants showed stable turn-taking behavior when playing music. The results of automated emotion classification using Support Vector Machines illustrates that emotional arousal in the ASD group can be detected and well recognized via EDA bio-signals. In summary, the results of our data analyses, including emotion classification using EDA signals, indicate that the proposed robot-music based therapy platform is an attractive and promising assistive tool to facilitate the improvement of fine motor control and turn-taking skills in children with ASD.
		
		\tiny
		\keyFont{ \section{Keywords:} Social Robotics, Autism, Music Therapy, Turn-Taking, Motor Control, Emotion Classification} 
	\end{abstract}
	
	\section{Introduction}
	\ Throughout history, music has been used medicinally due to its notable impact on the mental and physical health of its listeners. This practice became so popular over time that it ultimately transitioned into its own type of therapy called music therapy. The interactive nature of music therapy has also made it especially well-suited for children. Children's music therapy is performed either in a one-on-one session or a group session and by listening, singing, playing instruments, and moving, patients can acquire new knowledge and skills in a meaningful way. It has previously been shown to help children with communication, attention, and motivation problems as well as behavioral issues \cite{cibrian2020supporting,gifford2011using}.
	While music therapy is extremely versatile and has a broad range of applications, previous research studies have found  that using music as an assistive method is especially useful for children with autism \cite{cibrian2020supporting,mossler2019therapeutic, dellatan2003use,brownell2002musically,starr1998understanding}. This interest in music therapy as a treatment for children with autism has even dated as far back as the early $20^{th}$ century. In fact, in the 1940s, music therapy was used at psychiatric hospitals, institutions, and even schools for children with autism. Due to the fact both the diagnostic criteria for autism and music therapy as a profession were only just emerging at this time, no official documentation or research about this early work exists. In the next decade, the apparent unusual musical abilities of children with autism intrigued many music therapists \cite{warwick1991music}. By the end of the 1960s, music therapists started delineating goals and objectives for autism therapy \cite{reschke2011history}. Since the beginning of the 1970s and onward,  theoretically grounded music therapists have been working toward a more clearly defined approach to improve the lives of individuals with autism spectrum disorders (ASD). However, for decades, music therapists have not used a consistent assessment method when working with individuals with ASD. A lack of a quality universal assessment tool has caused therapy to drift towards non-goal driven treatment and has reduced the capacity to study the efficacy of the treatment in a scientific setting \cite{thaut2000scientific}.
	
	Humanoid socially assistive robots are ideal to address some of these concerns \cite{scassellati2012robots, diehl2012clinical}. Precise programming can be implemented to ensure the robots deliver therapy in a consistent assessment method each session. Social robots are also unique in their prior success in working with children with ASD \cite{boucenna2014interactive, pennisi2016autism}. Previous studies have found children with autism have less interest in communicating with humans due to how easily these interactions can become overwhelming \cite{marinoiu20183d,di2018deep,richardson2018robot,feng2013can} and instead are more willing to interact with humanoid social robots in daily life due to their relatively still faces and less intimidating characteristics. Some research has found that children with ASD speak more while interacting with a non-humanoid robot compared to regular human-human interactions \cite{kim2013social}. A ball-like robot called Sphero has been used to examine different play patterns and emotional responses of children with ASD \cite{boccanfuso2016emotional, boucenna2014learning}. Joint attention with body movement has also been tested and evaluated using a robot called NAO from \cite{anzalone2014children}. Music interaction has been recently introduced in Human-Robot-Interaction (HRI) as well \cite{taheri2019teaching}. The NAO robot has been widely used in this field such as music and dance imitation \cite{beer2016robot} and body gesture imitation \cite{guedjou2017influence,zheng2015robot,boucenna2016robots}.
	To expand upon the aforementioned implications of socially assistive robotics and music, \textbf{the scientific question of this research is whether music-therapy delivered by a socially assistive robot is engaging and effective enough to serve as a viable treatment option for children with ASD}.
	The main contribution of this paper is as follows. First, we propose a fully autonomous assistive robot-based music intervention platform for children with autism. Second, we designed HRI sessions and conducted studies based on the current platform, where motor control and turn-taking skills were practiced by a group of children with ASD. Third, we utilized and integrated machine learning techniques into our music-therapy HRI sessions to recognize and classify event-based emotion expressed by the study participants.
	
	The remainder of this paper is organized as follows. Section 2 presents related works concerning human-robot interaction in multiple intervention methods. Section 3 elaborates on the experiment design process including the details of the hardware and HRI sessions. We present our music therapy platform in Section 4 and the experimental results in Section 5. Finally, Section 6 concludes the paper with remarks for future work.
	
	\section{Related Works}
	
	\ Music is an effective method to involve children with autism in rhythmic and non-verbal communication and has often been used in therapeutic sessions with children who suffer from mental and behavioral disabilities \cite{lagasse2019assessing,boso2007effect,roper2003melodic}. Nowadays, at least 12\% of all treatments for individuals with autism consist of music-based therapies \cite{bhat2013review}. Specifically, playing music to children with ASD in therapy sessions has shown a positive impact on improving social communication skills \cite{lagasse2019assessing,lim2011effects}. Many studies have utilized both recorded and live music in interventional sessions for single and multiple participants \cite{dvir2020body,bhat2013review, corbett2008brief}. Different social skills have been targeted and reported (i.e., eye-gaze attention, joint attention and turn-taking activities) in music-based therapy sessions \cite{stephens2008spontaneous, kim2008effects}. Improving gross and fine motor skills for children with ASD through music interventions is  noticeably absent in this field of studies \cite{bhat2013review} and thus is one of the core features in the proposed study.
	Socially assistive robots are becoming increasingly popular as interventions for youth with autism. Previous studies have focused on eye contact and joint attention \cite{mihalache2020perceiving,  mavadati2014comparing,feng2013can}, showing that the pattern of gaze perception in the ASD group is similar to Typically Developing (TD) children as well as the fact eye contact skills can be significantly improved after interventional sessions. These findings also provide strong evidence that ASD children are more inclined to engage with humanoid robots in various types of social activities, especially if the robots are socially intelligent \cite{anzalone2015evaluating}. Other researchers have begun to use robots to conduct music-based therapy sessions. In such studies, children with autism are asked to imitate music based on the \textit{Wizard of Oz} and Applied Behavior Analysis (ABA) models using humanoid robots in interventional sessions to practice eye-gaze and joint attention skills \cite{askari2018pilot,taheri2015impact, taheri2016social}.
	
	However, past research has often had certain disadvantages, such as a lack of an automated system in human-robot interaction. This research attempts to address these shortcomings with our proposed platform. In addition, music can be used as a unique window into the world of autism as a growing body of evidence suggests that individuals with ASD are able to understand simple and complex emotions in childhood using music-based therapy sessions \cite{molnar2012music}. Despite the obvious advantages of using music-therapy to study emotion comprehension in children with ASD, limited research has been found, especially studies utilizing physiological signals for emotion recognition in ASD and TD children \cite{feng2018wavelet}. The current study attempted to address this gap by using electrodermal activity (EDA) as a measure of emotional arousal to better understand and explore the relationship between activities and emotion changes in children with ASD. To this end, the current research presents an automated music-based social robot platform with an activity-based emotion recognition system. The purpose of this platform is to provide a possible solution for assisting children with autism and help improve their motor and turn-taking skills. Furthermore, by using bio-signals with Complex-Morlet (C-Morlet) wavelet feature extraction \cite{feng2018wavelet}, emotion classification and emotion fluctuation can be analyzed based on different activities. TD children are included as a control group to compare the intervention results with the ASD group.
	
	\section{Experiment Design}
	\subsection{Participants Selection}
	Nine high functioning ASD participants (average age: 11.73, std: 3.11) and seven TD participants (average age: 10.22, std: 2.06) were recruited for this study. Only one girl was included in the ASD group, which according to \cite{loomes2017male}, is under the estimated ratio of 3:1 for male and females with ASD. ASD participants who had participated in previous unrelated research studies in the past were invited to return for this study. Participants were selected from this pool and the community with help from the University of Denver Psychology department. Six out of nine ASD participants had previous human-robot-interaction with another social robot named ZENO \cite{mihalache2020perceiving}, and three out of the nine kids from the ASD group had previous music experience on instruments other than the Xylophone (saxophone and violin). All children in the ASD group had a previous diagnosis of ASD in accordance with diagnostic criteria outlined in \cite{DSMIV2000}, including an ADOS report on record. Additionally, their parents completed the Social Responsiveness Scale (SRS) \cite{constantino2012social} for their child. The SRS provides a quantitative measure of traits associated with autism among children and adolescents between four and eighteen years-of-age. All ARS T-scores in our sample were above 65. Unlike the ASD group, TD kids had no experience with NAO or other robots prior to this study. Most of them, on the other hand, did participate in music lessons previously. Experimental protocols were approved by the University of Denver Institutional Review Board.
	
	\subsection{NAO: A Humanoid Robot}
	NAO, a humanoid robot merchandised by SoftBank Group Corporation, was selected for the current research. NAO is 58 cm (23 inches) tall and has 25 degrees of freedom, allowing him to perform most human body movements. According to the official Aldebaran manufacturer documentation, NAO's microphones have a sensitivity of 20mV/Pa +/-3dB at 1kHz and an input frequency range of 150Hz - 12kHz. Data is recorded as a 16 bit, 48000Hz, 4 channel wave file which meets the requirements for designing the online feedback audio score system. NAO’s computer vision module includes face and shape recognition units. By using the vision feature of the robot, NAO can see an instrument with its lower camera and implement an eye-hand self-initiative system that allows the robot to micro-adjust the coordination of its arm joints in real-time, which is very useful to correct cases of inproper positioning prior to engaging with the Xylophone.
	
	NOA’s arms have a length of approximately 31 cm. Position feedback sensors are equipped in each of the robot’s joints to obtain real-time localization information. Each robot arm has five degrees of freedom and is equipped with sensors to measure the position of joint movement. To determine the position of the Xylophone and the mallets’ heads, the robot analyzed images from the lower monocular camera located in its head, which has a diagonal field of view of 73 degrees. By using these dimensions, properly sized instruments can be selected and more accessories can be built.
	
	\subsection{Hardware Accessories}
	In order to have a well-functioning toy-size humanoid robot play music for children with autism, some necessary accessories needed to be made before the robot was capable of completing this task. All accessories will be discussed in the following paragraphs.
	
	In this system, due to the length of NAO’s outstretched arms, a Sonor Toy Sound SM Soprano Glockenspiel with 11 sound bars 2cm in width was selected and utilized in this research. The instrument is $31cm \times 9.5cm \times 4cm$, including the resonating body.
	The smallest sound bar is playable in an area of $2.8cm \times 2cm$, the largest in an area of $4.8cm \times 2cm$. The instrument is diatonically tuned in C major/A minor. The 11 bars of the Xylophone represent 11 different notes, or frequencies, that cover one and a half octave scales, from C6 to F7.
	The Xylophone, also known as the marimba or the glockenspiel, is categorized as a percussion instrument that consists of a set of metal/wooden bars that are struck with mallets to produce delicate musical tones. Much like the keyboard or drums, to play the Xylophone properly, a unique and specific technique needs to be applied. A precise striking movement is required to produce a beautiful note, an action perfect for practicing motor control, and the melody played by the user can support learning emotions through music.
	
	For the mallets, we used the pair that came with the Xylophone, but added a modified 3D-printed gripper that allowed the robot hands to hold them properly (see Figure \ref{griper}). The mallets are approximately 21 cm in length and include a head with a 0.8 cm radius. Compared to other designs, the mallet gripper we added encourages a natural holding position and allows the robot to properly model how participants should hold the mallet stick.
	\begin{figure}[tbp]
		\begin{center}
			\begin{tabular}{c}
				\epsfig{figure=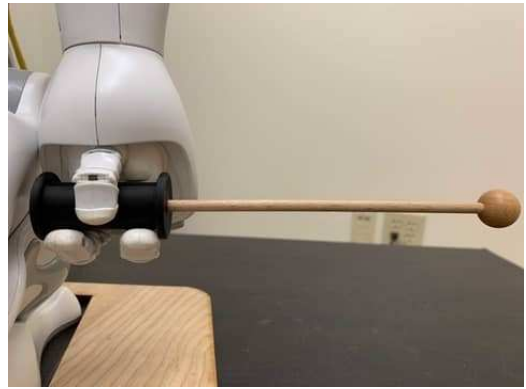, scale = 0.4}\label{griper} \\
			\end{tabular}
			\caption{Mallet Griper} \label{griper}
		\end{center}
	\end{figure}
	Using carefully measured dimensions, a wooden base was designed and laser cut to hold the Xylophone at the proper height for the robot to play in a crouching position. In this position, the robot could easily be fixed in a location and have the same height as the participants, making it more natural for the robot to teach activities (see Figure \ref{stand}).\\
	\begin{figure}[tbp]
		\begin{center}
			\begin{tabular}{c}
				\epsfig{figure=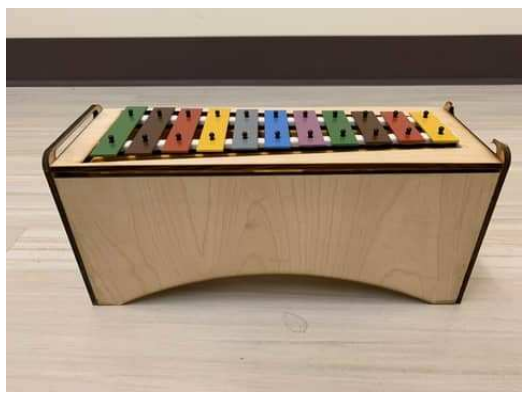, scale = 0.4} \label{front}
			\end{tabular}
			\caption{Instrument Stand Front View.} \label{stand}
		\end{center}
	\end{figure}
	\subsection{Q-Sensor}
	One Q-sensor \cite{kappas2013validation} was used in this study. Participants were required to wear this device during each session.  EDA signal
	(frequency rate 32Hz) was collected from the Q-sensor attached to the wrists (wrist side varied and was determined by the participants). Due to the fact participants often required breaks during sessions, 2 to 3 seperated EDA files were recorded. These files were annotated by comparing the time stamps with the videos manually.
	
	\subsection{Experiment Room Setup}
	All experiment sessions were held in an $11ft \times 9.5ft \times 10ft$ room with six HD surveillance cameras installed in the corners, on the sidewalls, and on the ceiling of the room (see Figure \ref{room}).The observation room was located behind a one-way mirror and participants were positioned with their backs toward this portion of the room to avoid distractions. During experiment sessions, an external, hand-held audio recorder was set in front of participants to collect high-quality audio to use for future research.
	\begin{figure*}[tbp]
		\begin{center}
			\begin{tabular}{c}
				\epsfig{figure=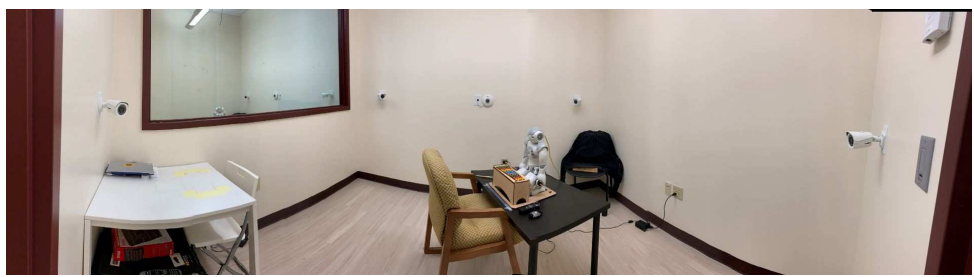, scale = .25}\label{room} 
			\end{tabular}
			\caption{Experiment room.} \label{room}
		\end{center}
	\end{figure*}
	 Utilizing a miniature microphone attatched to the ceiling camera, real-time video and audio were broadcasted to the observation room during sessions, allowing researchers to observe and record throughout the sessions. Parents or caregivers in the observation room could also watch and call off sessions in the case of an emergency.
	 
	\subsection{Experiment Sessions}
	For each participant in the ASD group, six total sessions were delivered, including one baseline session, four interventional sessions, and one exit session. Only the baseline and exit sessions were required for the TD group. Baseline and exit sessions contained two activities: 1) music practice and 2) music gameplay. Intervention sessions contained three parts: S1) warm-up; S2) single activity practice (with a color hint); and S3) music gameplay. Every session lasted for a total of 30-60 minutes depending on the difficulty of each session and the performance of individuals. Typically, each participant had baseline and exit session lengths that were comparable. For intervention sessions, the duration gradually increased in accordance with the increasing difficulty of subsequent sessions. Additionally, a user-customized song was used in each interventional session to have participants involved in multiple repetitive activities. The single activity practice was based on music practice from the baseline/exit session. In each interventional session, the single activity practice only had one type of music practice. For instance, single-note play was delivered in the first intervention session. For the next intervention session, the single activity practice increased in difficulty to multiple notes. The level of difficulty for music play was gradually increased across sessions and all music activities were designed to elicit an emotional reaction.
	
\begin{figure*}[tbp]
		\begin{center}
			\begin{center}
				\epsfig{figure=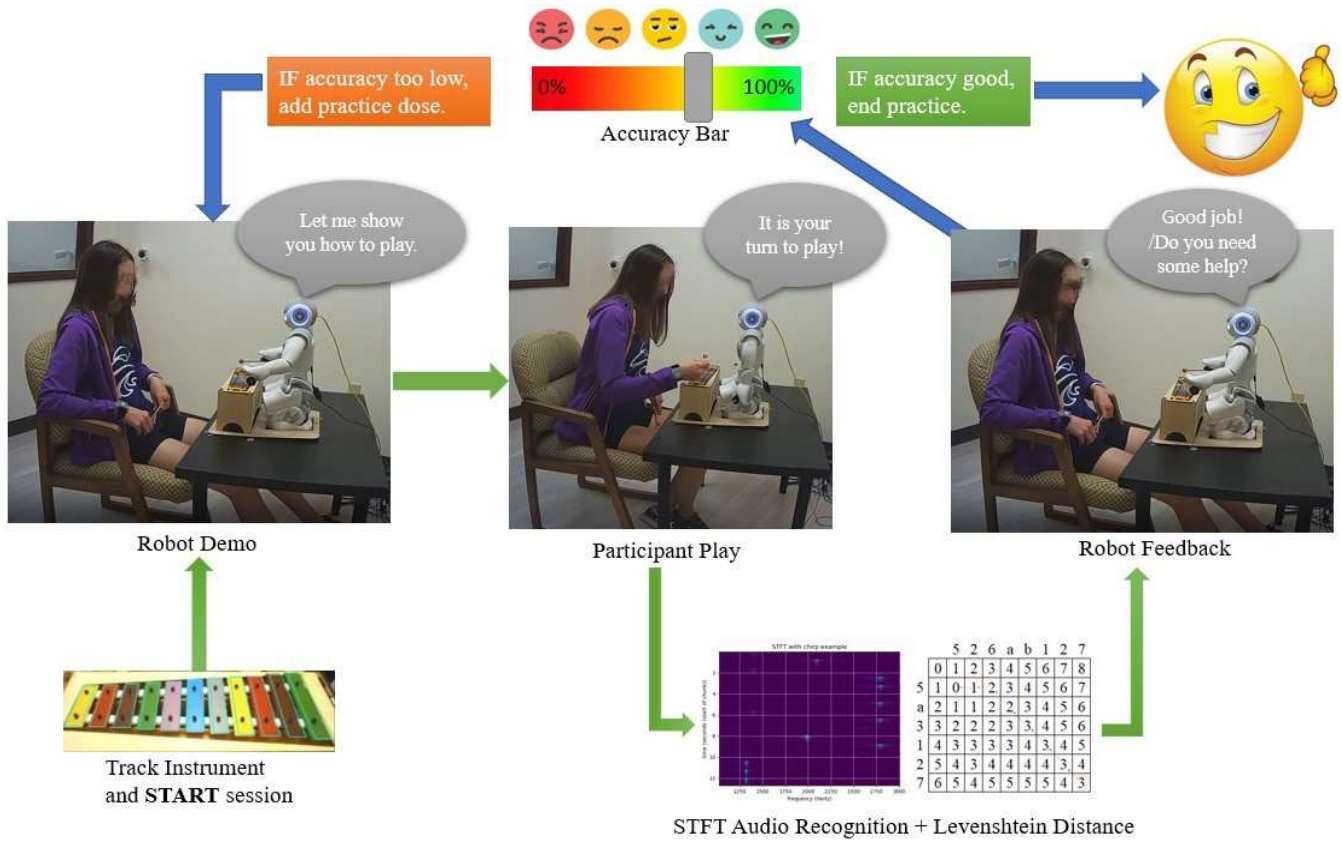, scale = .45}\label{r} 
			\end{center}
			\caption{Experiment session illustration.} \label{interact}
		\end{center}
	\end{figure*}

	To aid in desired music-based social interaction, NAO delivered verbal and visual cues indicating when the participant should take action. A verbal cue consisted of the robot prompting the participant with “Now, you shall play right after my eye flashes.” Participants could also reference NAO's eyes changing color as a cue to start playing. Aftter the cues, participants were then given five to 10 seconds to replicate NAO's strikes on the Xylophone. The start of each session was considered a warmup and participants were allowed to imitate NAO freely without accuracy feedback. The purpose of having a warm-up section was to have participants focus on practicing motor control skills while also refreshing their memory on previous activities. Following the warmup, structured activities would begin with the goal of the participant ultimately completing a full song. Different from the warmup section, notes played during this section in the correct sequence were considered to be a good-count strike and notes incorrectly played were tracked for feedback and additional training purposes. To learn a song, NAO gradually introduced elements of the song, starting with a single note and color hint. Subsequently, further notes were introduced, then the participants were asked to play half the song, and then finally play the full song after they perfected all previous tasks. Once the music practice was completed, a freshly designed music game containing three novel entertaining game modes was presented to the participants. Participants could then communicate with the robot regarding which mode to play with. Figure \ref{interact} shows the complete interaction process in our HRI study.
	
	The laboratory where these experiments took place is 300-square feet and has environmental controls that allowed the temperature and humidity of the testing area to be kept consistent. For experimental purposes in the laboratory, the ambient temperature and humidity were kept constant to guarantee the proper functionality of the Q-sensor as well as the comfort of subjects in light clothing. To obtain appropriate results and ensure the best possible function of the device, the sensor was cleaned before each usage. Annotators were designated to determine the temporal relation between the video frames and the recorded EDA sequences of every participant. To do this, annotators went through video files of every session, frame by frame, and designated the initiation and conclusion of an emotion. Corresponding sequences of EDA signals were then identified and utilized to create a dataset for every perceived emotion. 
	
	Figure \ref{eda_anno} shows the above-described procedure diagrammatically. Due the fact that it is hard to conclude these emotions with specific facial expressions, event numbers will be used in the following analysis section representing emotion comparison.  The rest of the emotion labels respresent the dominant feeling of each music activity according to the observations of both annotators and the research assistant who ran the experiment sessions. 
	\begin{figure}[tbp]
		\begin{center}
			\begin{center}
				\includegraphics[width=1\linewidth]{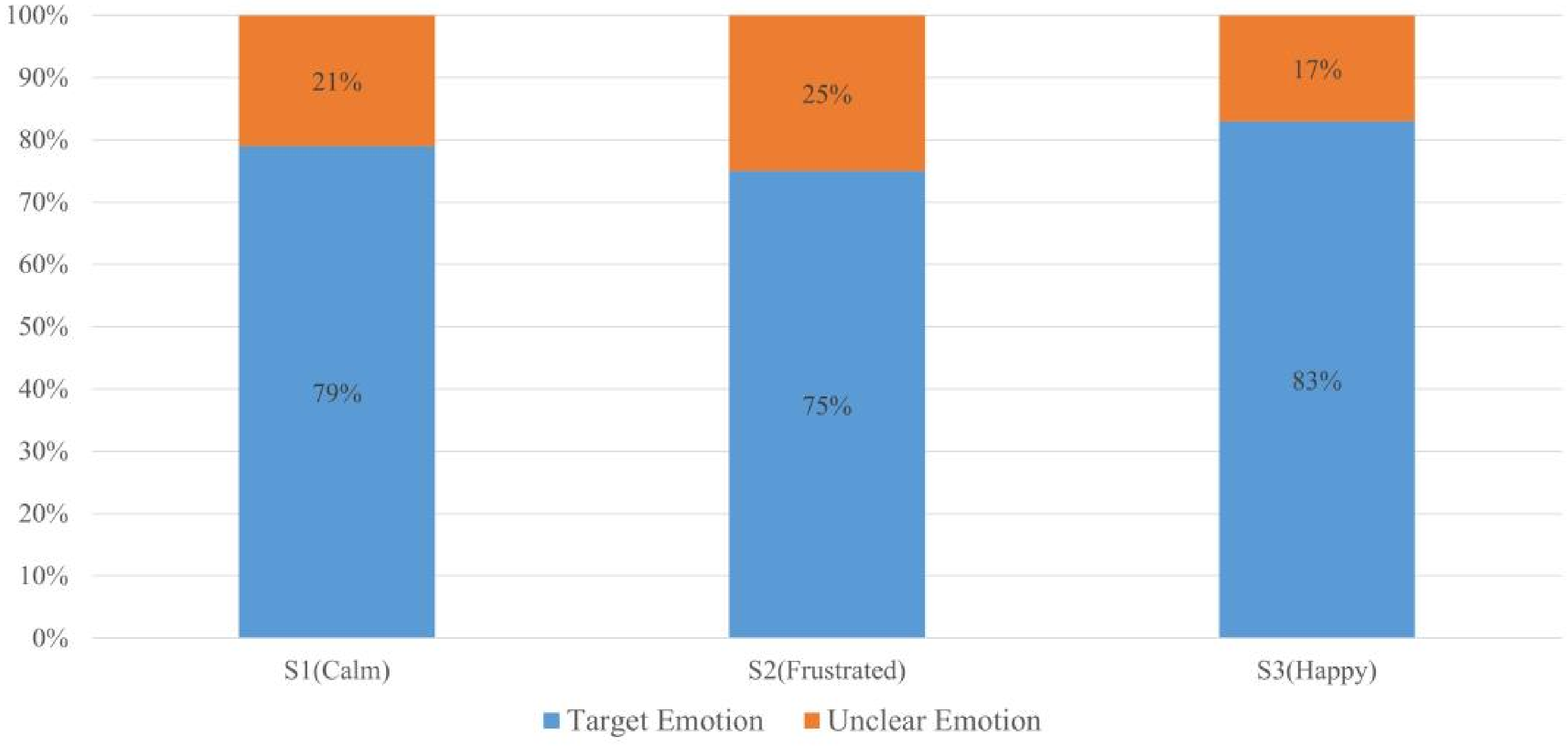}\label{eda_anno}
			\end{center}
			\caption{The distribution of the targeted emotions across all subjects and events. 
			} \label{eda_anno}
		\end{center}
	\end{figure}

	\;\noindent\textbf{Mode 1):} The robot randomly picks a song from its song bank and plays it for participants. After each song, participants were asked to identify a feeling they ascribed to the music to find out whether music emotion can be recognized. 
	
	\textbf{Mode 2):} A sequence of melodies were randomly generated by the robot with a consonance (happy or comfortable feeling) or dissonance (sad or uncomfortable feeling) style. Participants were asked to express their emotions verbally and a playback was required afterwards.
	
	\textbf{Mode 3):} Participants have five seconds of free play then challenge the robot to imitate what was just played. After the robot was done playing, the participant rated the performance, providing a reversed therapy-like experience for all human volunteers. 
	
	There was no limit on how many trials or modes each individual could play for each session, but each mode had to be played at least once in a single session. The only difference between the baseline and exit sessions was the song that was used in them. In the baseline session, "Twinkle, Twinkle, Little Star" was used as an entry-level song for every participant. For the exit session, participants were allowed to select a song of their choosing so that they would be more motivated to learn the music. In turn, this made the exit session more difficult than the baseline session. By using the Module-Based Acoustic Music Interactive System, inputting multiple songs became possible and less time-consuming. More than 10 songs were collected in the song bank including "Can Can" by Offenbach, "Shake It Off" by Taylor Swift, the "SpongeBob SquarePants" theme from the animated show SpongeBob SquarePants, and "You Are My Sunshine" by Johnny Cash. Music styles covered kid’s songs, classic, pop, ACG (Anime, Comic and Games), and folk, highlighting the versatile nature of this platform. In addition to increasing the flexibility of the platform, having the ability to utilize varying music styles allows the robot to accommodate a diverse range of personal preferences, further motivating users to learn and improve their performance.
	
	\section{Module-Based Acoustic Music Interactive System Design}
	In this section, a novel, module-based robot-music therapy system will be presented. For this system to be successful, several tasks had to be accomplished: a) allow the robot to play a sequence of notes or melody fluently; b) allow the robot to play notes accurately; c) allow the robot to adapt to multiple songs easily; d) allow the robot to be able to have social communication with participants; e) allow the robot to be able to deliver learning and therapy experiences to participants; and f) allow the robot to have fast responses and accurate decision-making. To accomplish these tasks, a module-based acoustic music interactive system was designed. Three modules were built in this intelligent system: Module 1: eye-hand self-calibration micro-adjustment; Module 2: joint trajectory generator; and Module 3: real-time performance scoring feedback (see Figure \ref{module}).
	
	\begin{figure}
		\begin{center}
			\begin{tabular}{c}
				\includegraphics[width=0.9\linewidth]{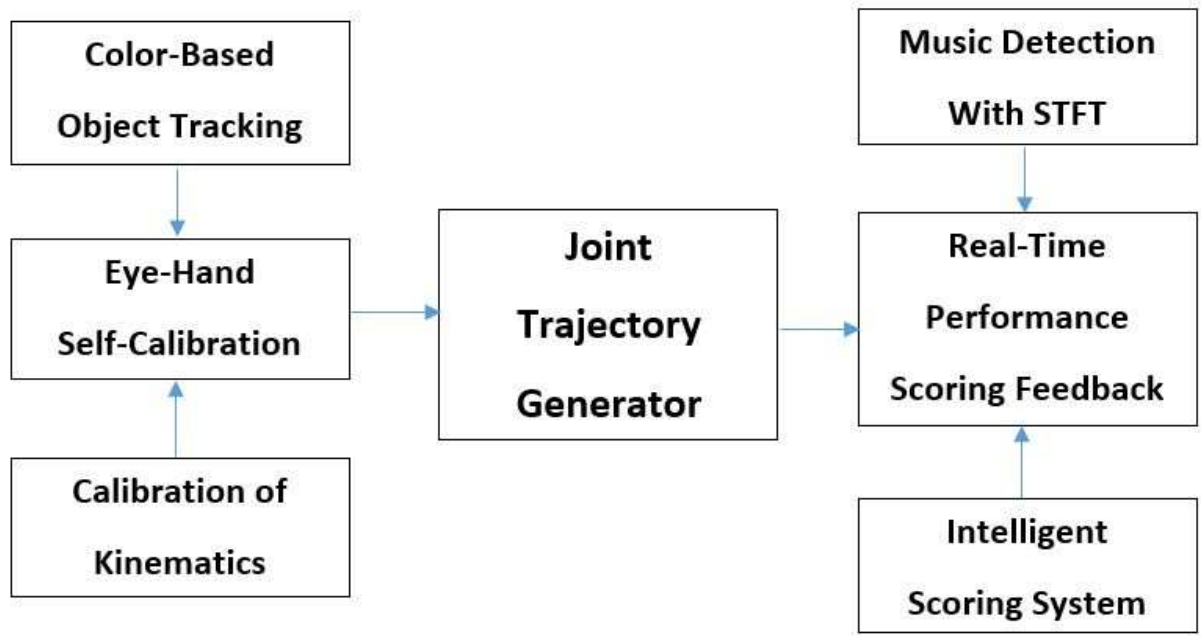}\label{module}\\
			\end{tabular}
			\caption{Block diagram of Module-based acoustic music interactive system.} \label{module}
		\end{center}
	\end{figure}
	\subsection{Module 1: Color-Based Position Initiative}
	Knowledge about the parameters of the robot’s kinematic model was essential for programming tasks requiring high precision, such as playing the Xylophone. While the kinematic structure was known because of the construction plan, errors could occur because of factors such as imperfect manufacturing. After multiple rounds of testing, it was determined that the targeted angle chain of arms never equals the returned chain. Therefore, we used a calibration method to eliminate this error.
	
	To play the Xylophone, the robot had to be able to adjust its motions according to the estimated relative position of the Xylophone and the heads of the mallets it was holding. To estimate the poses adopted in this paper, we used a color-based technique.
	
\begin{figure}
		\begin{center}
			\begin{tabular}{c}
				\includegraphics[width=0.4\linewidth]{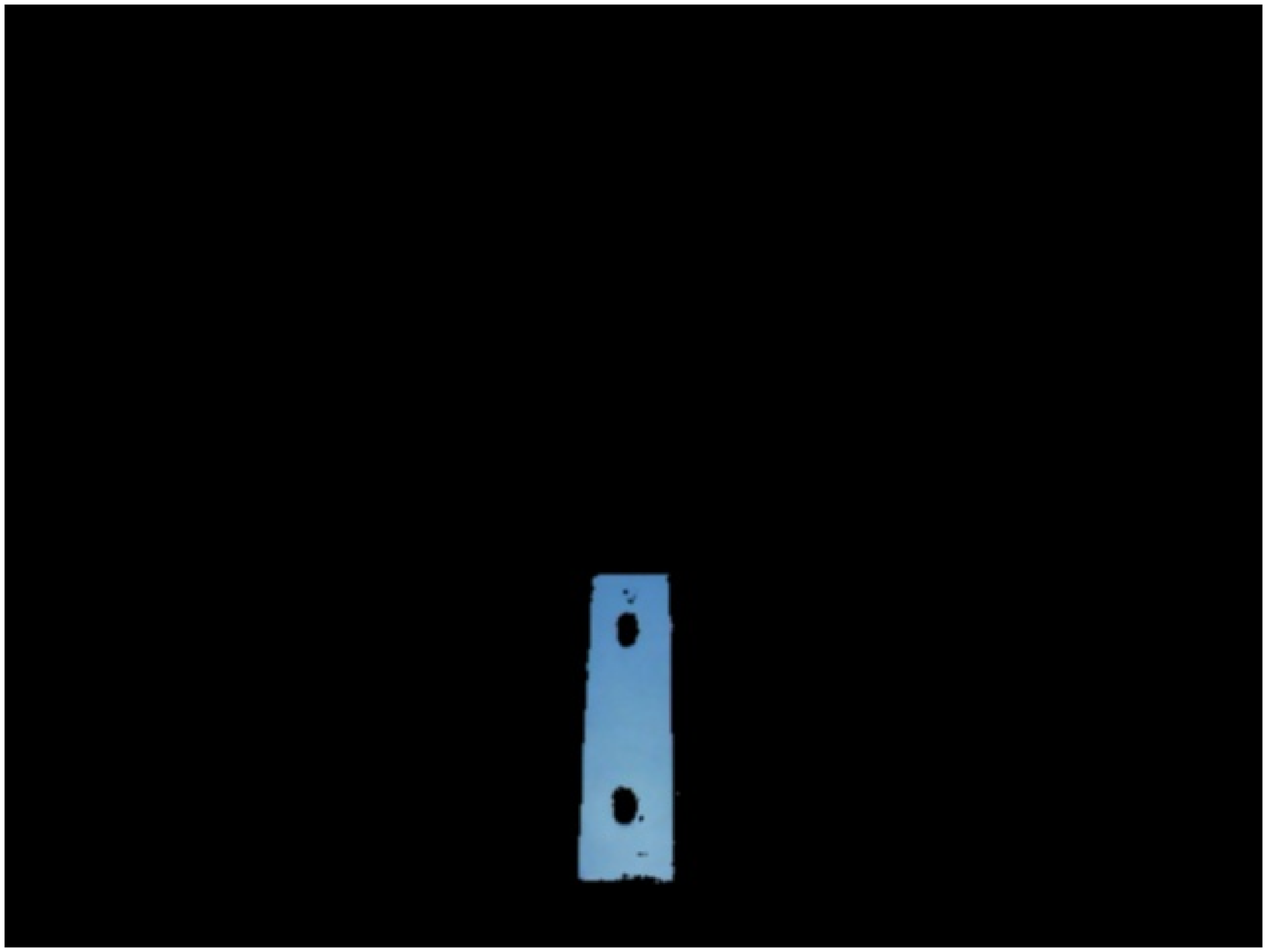}\label{single_color_a}\\
				(a)\\
				\includegraphics[width=0.4\linewidth]{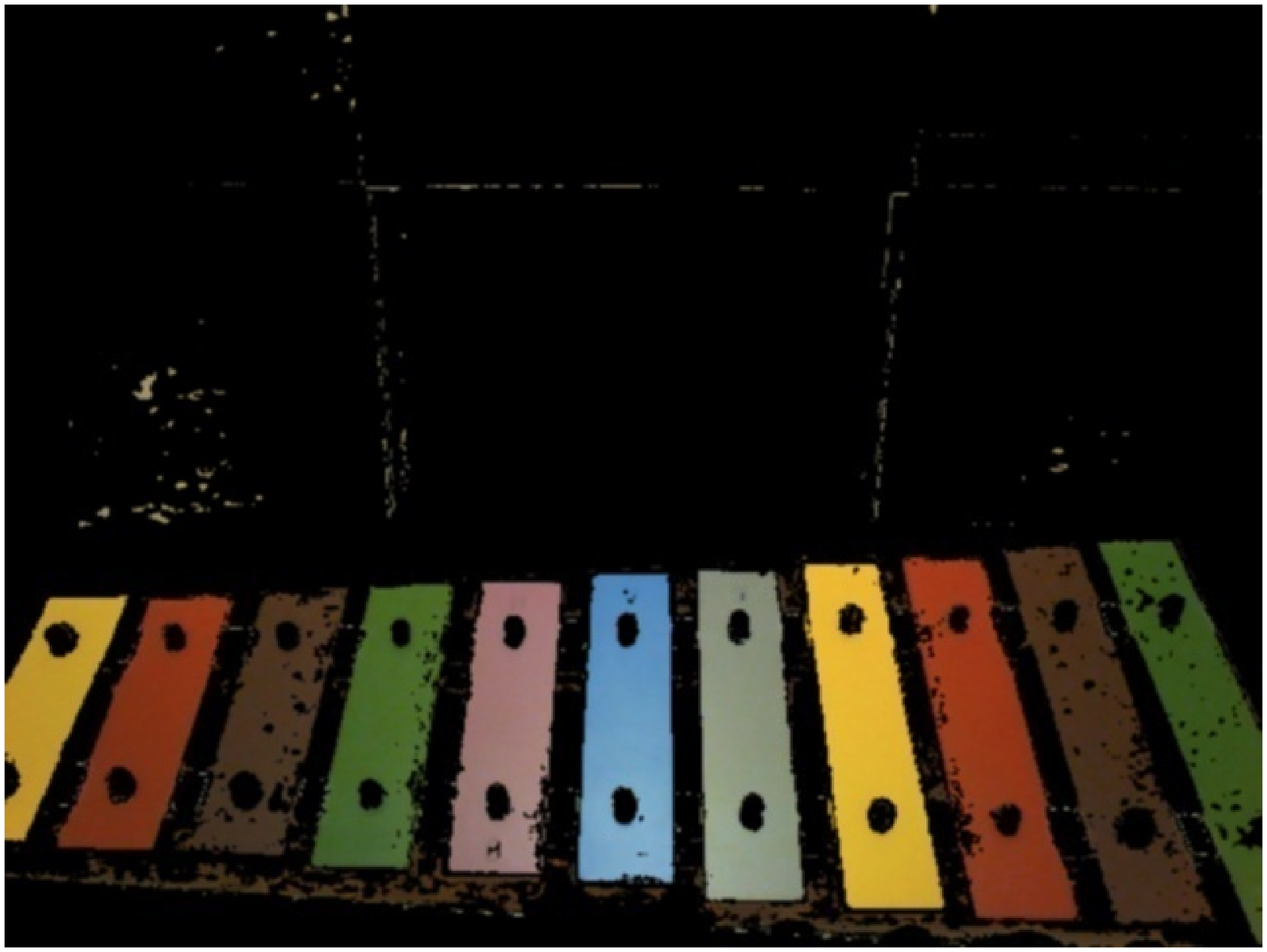}\label{single_color_b}\\
				(b)\\
				\includegraphics[width=0.75\linewidth]{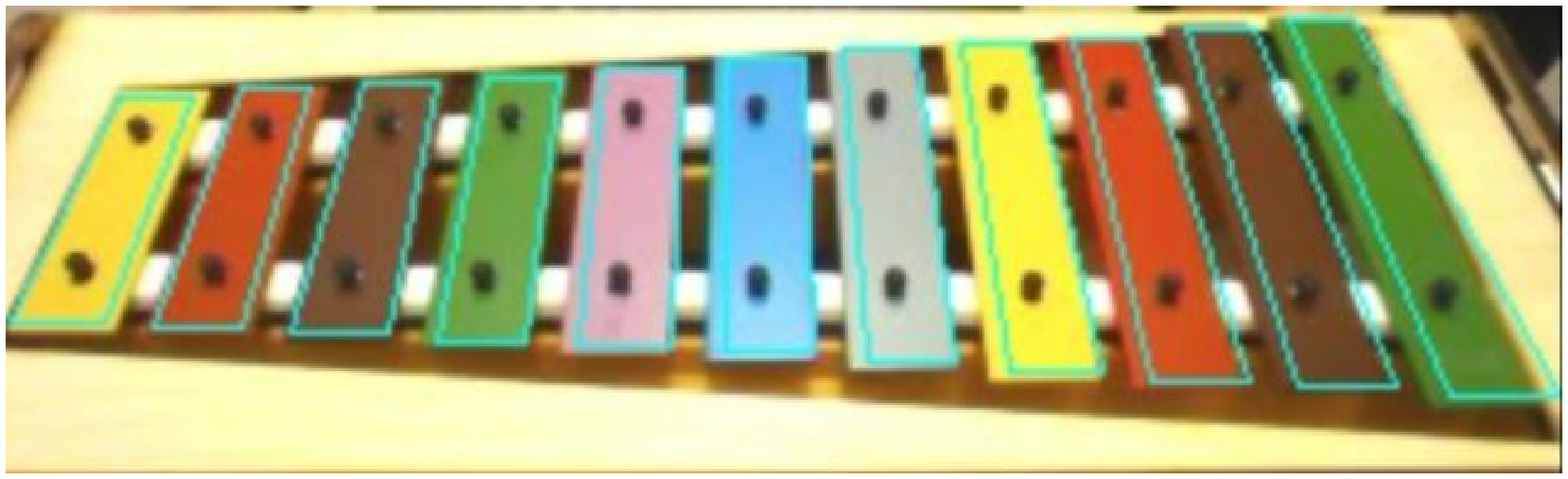}\label{color_detection_c}\\
				(c)
			\end{tabular}
			\caption{Color detection from NAO's bottom camera: (a) single blue color detection (b) full instrument color detection (c) color based edge detection.} \label{color_detection}
		\end{center}
	\end{figure}

	The main idea in object tracking is that, based on the RGB color of the center blue bar, given a hypothesis about the Xylophone’s position, one can project the contour of the Xylophone’s model into the camera image and compare them to an observed contour. In this way, it is possible to estimate the likelihood of the position hypothesis. Using this method, the robot can track the Xylophone with extremely low cost in real-time (see Figure \ref{color_detection}).
	
	\subsection{Module 2: Joint Trajectory Generator}
	Our system parsed a list of hexadecimal numbers (from 1 to b) to obtain the sequence of notes to play. The system converted the notes into a joint trajectory using the striking configurations obtained from inverse kinematics as control points. The timestamps for the control points are defined by the user to meet the experiment requirement. The trajectory was then computed by the manufacturer-provided API, using Bezier curve \cite{han2008novel} interpolation in the joint space, and then sent to the robot controller for execution. This process allowed the robot to play in-time with songs. 
	
	\subsection{Module 3: Real-Time Performance Scoring Feedback}
	Two core features were designed to complete the task in the proposed scoring system: 1) music detection and 2) intelligent scoring feedback. These two functions could provide a real-life interaction experience using a music therapy scenario to teach participants social skills.
	
	\subsubsection{Music Detection}
	Music, from a science and technology perspective, is a combination of time and frequency. To make the robot detect a sequence of frequencies, we adopted the Short-time Fourier transform (STFT) for its audio feedback system. Doing so allowed the robot to be able to understand the music played by users and provide proper feedback as a music instructor.
	
	The STFT is a Fourier-related transform used to determine the sinusoidal frequency and phase content of local sections of a signal as it changes over time. In practice, the procedure for computing STFTs is to divide a longer time signal into shorter segments of equal length and then separately compute the Fourier transform for each shorter segment. Doing so reveals the Fourier spectrum on each shorter segment. The changing spectra can then be plotted as a function of time. In the case of discrete-time, data to be transformed can be broken up into chunks of frames that usually overlap each other to reduce artifacts at boundaries. Each chunk is Fourier transformed. The complex results are then added to a matrix that records magnitude and phase for each point in time and frequency (see Figure \ref{stft}). This can be expressed as:
	\begin{equation}
		\resizebox{.6\hsize}{!}{${\displaystyle \mathbf {STFT} \{x[n]\}(m,\omega )\equiv X(m,\omega )=\sum _{n=-\infty }^{\infty }x[n]w[n-m]e^{-j\omega n}}$}
	\end{equation}
	\noindent likewise, with signal x[n] and window w[n]. In this case, m is discrete and $\omega$ is continuous, but in most applications, the STFT is performed on a computer using the Fast Fourier Transform, so both variables are discrete and quantized. The magnitude squared of the STFT yields the spectrogram representation of the Power Spectral Density of the function:
	\begin{equation}
		\resizebox{.4\hsize}{!}{${\displaystyle \operatorname {spectrogram} \{x(t)\}(\tau ,\omega )\equiv |X(\tau ,\omega )|^{2}}$}
	\end{equation}
	\noindent After the robot detects the notes from user input, a list of hexadecimal numbers are returned. This list is used for two purposes: 1) to compare with the target list for scoring and sending feedback to the user and 2) to create a new input for having robot playback in the game session.
	
	\begin{figure}
		\begin{center}
			\begin{tabular}{c}
				\includegraphics[width=1\linewidth]{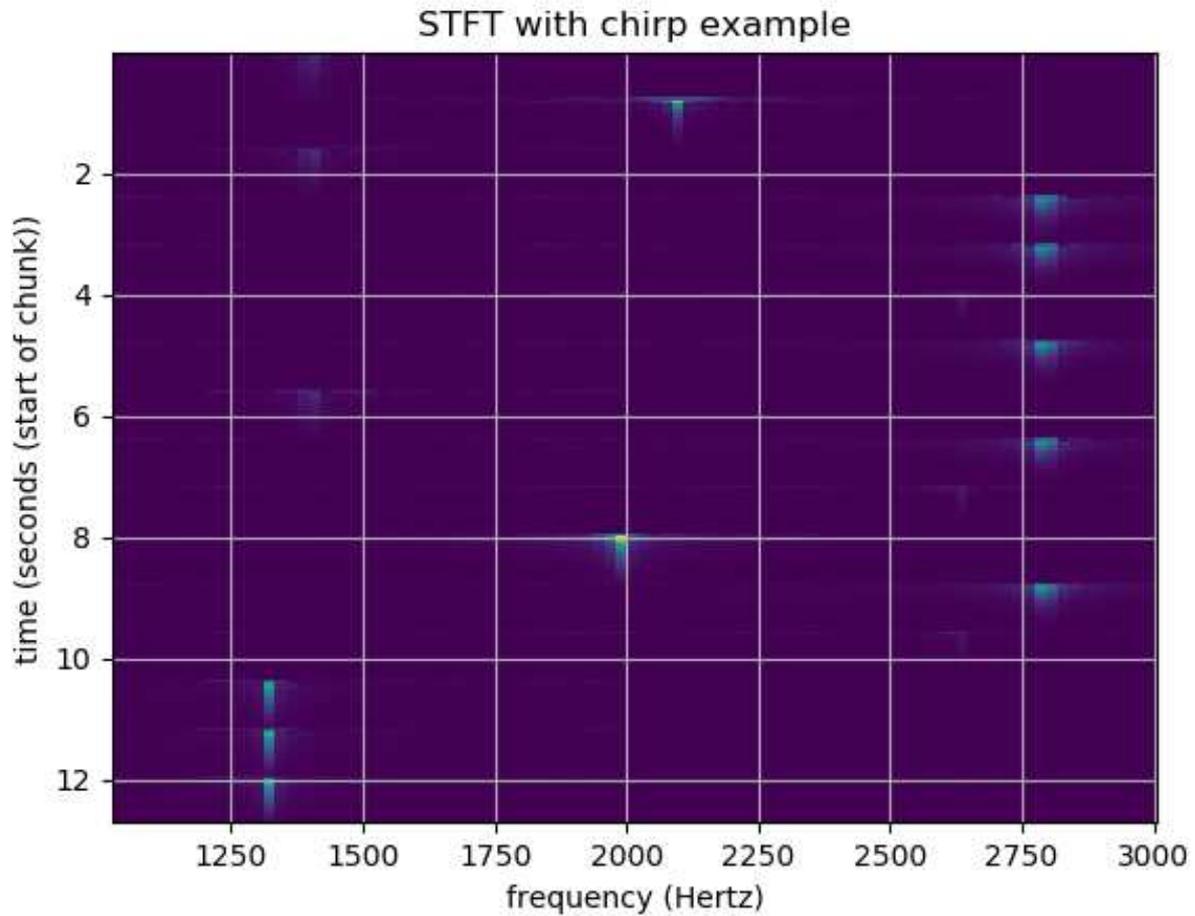}\label{stft}
			\end{tabular}
			\caption{Melody detection with Short Time Fourier Transform} \label{stft}
		\end{center}
	\end{figure}
	\subsubsection{Intelligent Scoring-Feedback System}
	To compare the detected and target notes, we used Levenshtein distance, an algorithm that is typically used in information theory linguistics. This algorithm is a string metric for measuring the difference between two sequences.
	
	In our case, the Levenshtein distance between two string-like hex-decimal numbers 
	${\displaystyle a,b}$ (of length ${\displaystyle |a|}$ and ${\displaystyle |b|}$ respectively) 
	is given by ${\displaystyle \operatorname {lev} _{a,b}(|a|,|b|)}$,
	\begin{equation}
		\resizebox{.8\hsize}{!}{${\displaystyle \qquad \operatorname {lev} _{a,b}(i,j)={\begin{cases}\max(i,j)&{\text{ if }}\min(i,j)=0,\\\min {\begin{cases}\operatorname {lev} _{a,b}(i-1,j)+1\\\operatorname {lev} _{a,b}(i,j-1)+1\\\operatorname {lev} _{a,b}(i-1,j-1)+1_{(a_{i}\neq b_{j})}\end{cases}}&{\text{ otherwise.}}\end{cases}}}$}\\
	\end{equation}
	\noindent where ${\displaystyle 1_{(a_{i}\neq b_{j})}}$ is the indicator function equal to 0 when 
	${\displaystyle a_{i}=b_{j}}$ and equal to 1 otherwise, and ${\displaystyle \operatorname {lev} _{a,b}(i,j)}$ 
	is the distance between the first ${\displaystyle i}$ characters of ${\displaystyle a}$ and the
	first ${\displaystyle j}$ characters of ${\displaystyle b}$.
	Note that the first element in the minimum corresponds to deletion (from ${\displaystyle a}$ to 
	${\displaystyle b}$), the second to insertion and the third to match or mismatch, depending on 
	whether the respective symbols are the same. 
	
	Based on the real-life situation, we defined a likelihood margin for determining whether the result
	is good or bad:
	\begin{equation}
		\resizebox{.45\hsize}{!}{${likelihood = \dfrac{len(target) - lev_{target,source}}{len(target)}}$}
	\end{equation}
	\noindent where, if the likelihood is more than 66\% (not including single note practice since, in that case, it is only correct or incorrect), the system will consider it to be a good result. This result is then passed to the accuracy calculation system to have the robot decide whether it needs to add additional practice trials (e.g., 6 correct out of a total of 10 trials).
	
	\section{Social Behavior Results}
	\subsection{Motor Control}
	Nine ASD and seven TD participants completed the study over the course of eight months. All ASD participants completed all six sessions (baseline, intervention, and exit) while all the TD subjects completed the required baseline and exit sessions. Conducting a Wizard of Oz experiment, a well-trained researcher was involved in the baseline and exit sessions in order for there to be high-quality observations and performance evaluations. With well-designed, fully automated intervention sessions, NAO was able to initiate music-based therapy activities with participants.
	
	Since the music detection method was sensitive to the audio input, a clear and long-lasting sound from the Xylophone was required. As seen in Figure \ref{fig9}, it is evident that most of the children were able to strike the Xylophone properly after one or two sessions (the average accuracy is improved over sessions). Notice that participants 101 and 102 had significant improvement during intervention sessions. Some of the participants started at a higher accuracy rate and sustained a rate above 80\% for the duration of the study. Even with some oscillation, participants with this type of accuracy rate were considered to have consistent motor control performance. 
	
	\begin{figure}[tbp]
		\begin{center}
			\begin{center}
				\includegraphics[width=1\linewidth]{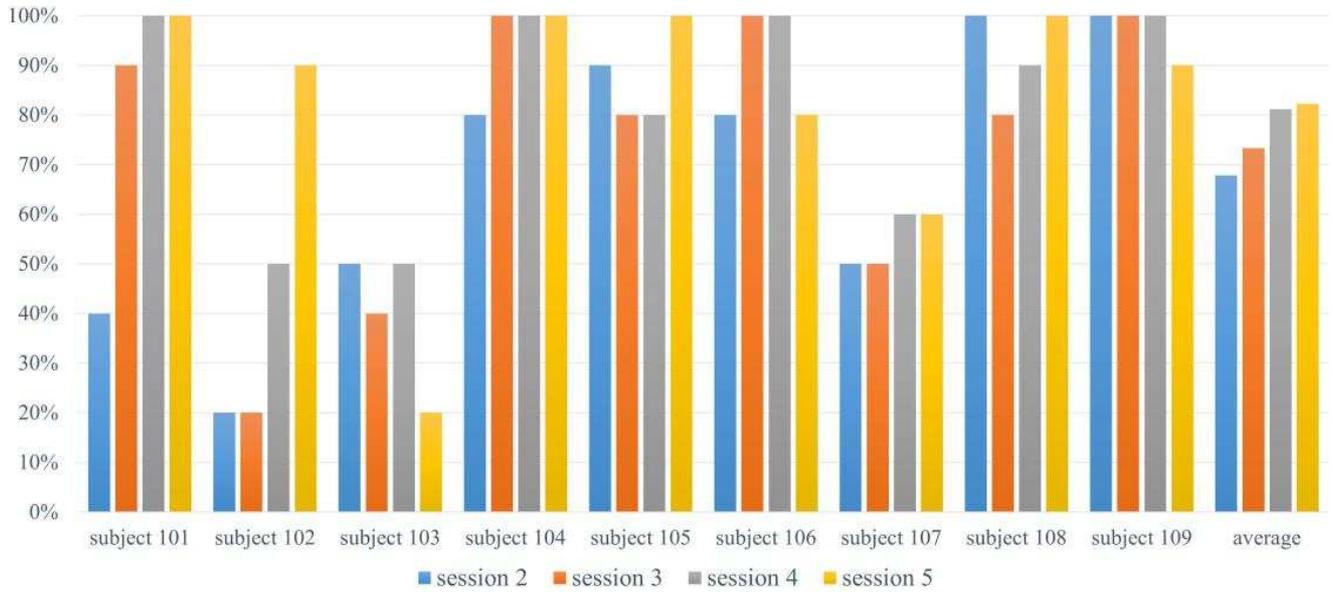}\label{fig9}\\
			\end{center}
			\caption{Motor control accuracy result.} \label{fig9}
		\end{center}
	\end{figure}
	\begin{figure}[tbp]
		\begin{center}
			\begin{center}
				\includegraphics[width=1\linewidth]{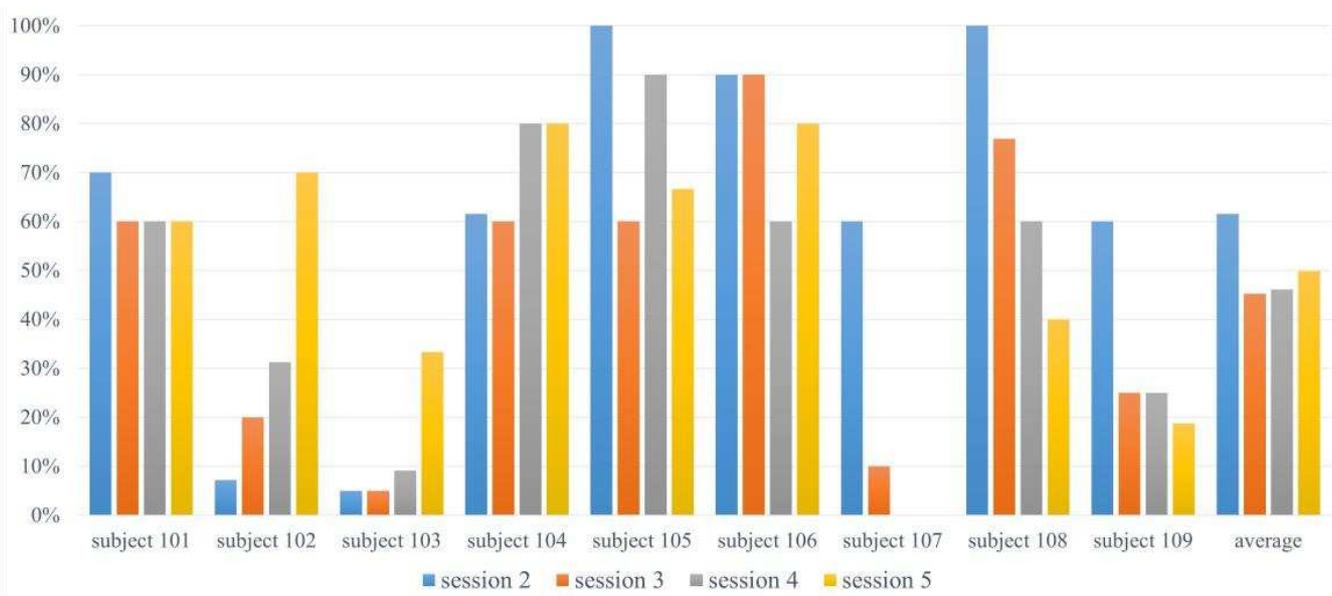}\label{fig10}\\
			\end{center}
			\caption{Main music therapy performance accuracy.} \label{fig10}
		\end{center}
	\end{figure}
	Figure \ref{fig10} shows the accuracy of the first music therapy activity that was part of the intervention sessions across all participants. As described in the previous section, the difficulty level of this activity was designed to increase across sessions. With this reasoning, the accuracy of participant performance was expected to decrease or remain consistent. This activity required participants to be able to concentrate, use joint attention skills during the robot therapy stage, and respond properly afterwards.  As seen in Figure \ref{fig10}, most of the participants were able to complete single/multiple notes practice with an average of a 77.36\%/69.38\% accuracy rate although, even with color hints, the pitch difference of notes was still a primary challenge. Due to the difficulty of sessions 4 and 5, a worse performance, when compared to the previous two sessions, was acceptable. However, more than half of the participants consistently showed a high accuracy performance or an improved performance than previous sessions. 
	
	\subsection{Turn-Taking Behavior}
    The American Psychological Association \cite{vandenbos2007american} defines turn-taking as, "in social interactions, alternating behavior between two or more individuals, such as the exchange of speaking turns between people in conversation or the back-and-forth grooming behavior that occurs among some nonhuman animals. Basic turn-taking skills are essential for effective communication and good interpersonal relations, and their development may be a focus of clinical intervention for children with certain disorders (e.g., Autism)." Learning how to play one’s favorite song can be a motivation that helps ASD participants understand and learn turn-taking skills. In our evaluation system, the annotators scored turn-taking based on how well kids were able to follow directions and display appropriate behavior during music-play interaction. In other words, participants needed to listen to the robot's instructions and then play, and vice versa. Any actions not following the interaction routine were considered as less or incomplete turn-taking. Measurements are described as follows: measuring turn-taking behavior can be subjective, so to quantify this, a grading system was designed. Four different behaviors were defined in the grading system: (a) “well-done”, this level is considered good behavior where the participant should be able to finish listening to the instructions from NAO, start playback after receiving the command, and wait for the result without interrupting (3 points for each “well-done”); (b) “light-interrupt”, in this level, the participant may exhibit slight impatience, such as not waiting for the proper moment to play or not paying attention to the result (2 points given in this level); (c) “heavy-interrupt”, more interruptions may accrue in this level and the participant may interrupt the conversation, but is still willing to play back to the robot (1 point for this level); and (d) “indifferent”, participant shows little interest in music activities including, but not limited to, not following behaviors, not being willing to play, not listening to the robot or playing irrelevant music or notes (score of 0 points). The higher the score, the better turn-taking behavior the participant demonstrated. All scores were normalized into percentages due to the difference of total “conversation” numbers. Figure \ref{fig11} shows the total results among all participants. Combining the report from video annotators, 6 out of 9 participants exhibited stable, positive behavior when playing music, especially after the first few sessions. Improved learn-and-play turn-taking rotation was demonstrated over time. Three participants demonstrated a significant increase in performance, suggesting turn-taking skills were taught in this activity. Note that two participants (103 \& 107) had a difficult time playing the Xylophone and following turn-taking cues given by the robot.
	
	\begin{figure}[tbp]
		\begin{center}
			\begin{center}
				\includegraphics[width=1\linewidth]{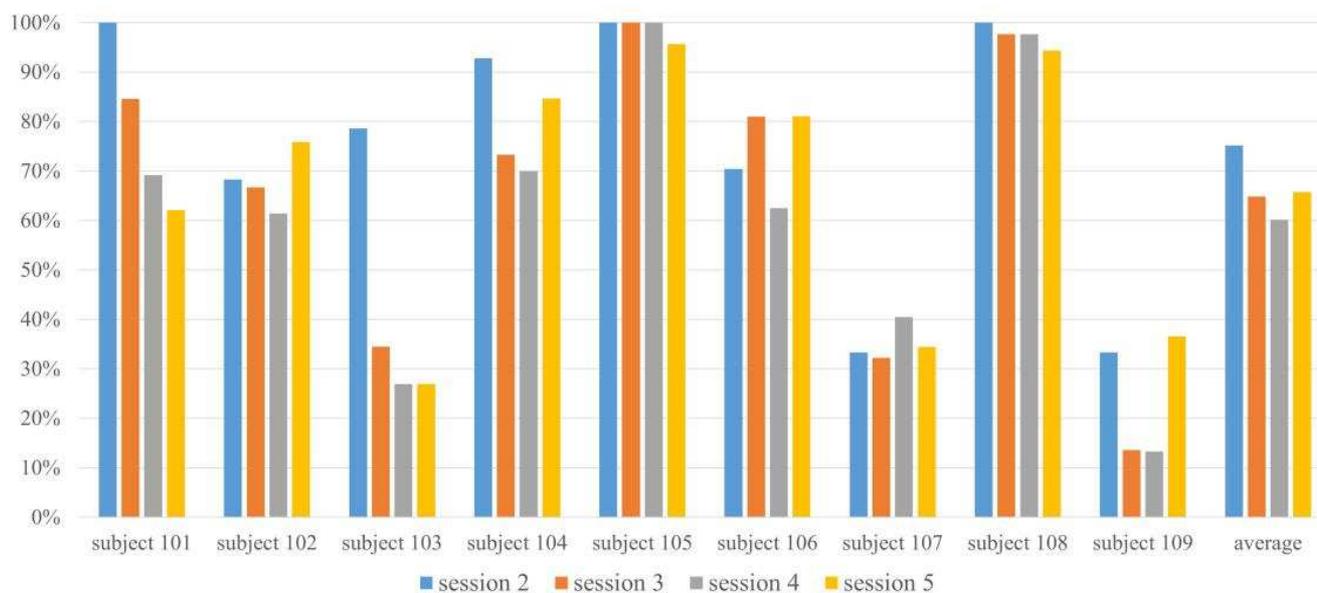}\label{fig11}\\
			\end{center}
			\caption{Normalized turn-taking behavior result for all subjects during intervention sessions.
			} \label{fig11}
		\end{center}
	\end{figure}
	
	\section{Music Emotion Classification Results}
	\ Since we developed our emotion classification method based on the time-frequency analysis of EDA signals, the main properties of The Continuous Wavelet Transforms (CWT), assuming that it is a C-Morlet wavelet, is presented here. Then, the pre-processing steps, as well as the wavelet-based feature extraction scheme, are discussed. Finally, we briefly review the characteristics of the SVM, the classifier used in our approach.
	
	EDA signals were collected in this study. By using the annotation and analysis method from our previous study \cite{feng2018wavelet}, we were able to produce a music-event-based emotion classification result that is presented below. To determine the emotions of the ASD group, multiple comparisons were made after annotating the videos. Note that on average, 21\% of emotions were not clear during the annotation stage. It is necessary to have them as unclear rather than label them with specific categories.
	
	During the first part of the annotation, it was not obvious to conclude the facial expression changes from different activities, however, in S1 participants showed more "calm" for most of the time as the dominant emotion due to the simplicity of completing the task. In S2, the annotator could not have a precise conclusion by reading the facial expression from participants as well. Most of the participants intended to play music and complete the task, however, feelings of "frustration" were often invoked in this group. A very similar feeling can be found in S3 when compared to S1. In S3, all the activities were designed to create a role changing environment for users in order to stimulate a different emotion. Most of the children showed "happy" emotion during the music game section.
	 We noticed that different activities can result in a change in emotional arousal. As mentioned above, the warm-up section and single activity practice section use the same activity with different intensities levels. The gameplay section has the lowest difficulty and is purposely designed to be more relaxing.
	\begin{table*}
		\label{tab1}
		\begin{center}
			\caption{Emotion change in different events using wavelet-based feature extraction under SVM classifier. }
			\vspace{3mm}
			\begin{tabular}{llllll}
				& Kernels                     & Accuracy(\%) & AUC & Precision(\%) & Recall(\%) \\
				\hline
				S1 vs S2       & \multirow{4}{*}{\textbf{Linear}}     & 75       & .78  & 76        & 72     \\
				S1 vs S3       &                             & 57       & .59  & 56        & 69     \\
				S2 vs S3       &                             & 69       & .72  & 64        & 86     \\
				S1 vs S2 vs S3 &                             & \multicolumn{4}{l}{52}              \\
				\hline
				S1 vs S2       & \multirow{4}{*}{\textbf{Polynomial}} & 66       & .70  & 70        & 54     \\
				S1 vs S3       &                             & 64       & .66  & 62        & 68     \\
				S2 vs S3       &                             & 65       & .68  & 62        & 79     \\
				S1 vs S2 vs S3 &                             & \multicolumn{4}{l}{50}              \\
				\hline
				S1 vs S2       & \multirow{4}{*}{\textbf{RBF}}        & 76       & .81  & 76        & 75     \\
				S1 vs S3       &                             & 57       & .62  & 57        & 69     \\
				S2 vs S3       &                             & 70       & .76  & 66        & 83     \\
				S1 vs S2 vs S3 &                             & \multicolumn{4}{l}{53}              \\
				\hline
			\end{tabular}
			\label{tab1}
		\end{center}
	\end{table*}
	
	\begin{table*}[tbp]
		\label{tab2}
		\begin{center}
			\caption{Emotion change classification performance in single event with segmentation using both SVM and KNN classifier. }
			\vspace{3mm}
			\resizebox{\columnwidth}{!}{
				\begin{tabular}{ccccccccc}
					\multicolumn{1}{l}{\multirow{3}{*}{}} & \multicolumn{5}{c}{Segmentation Comparison in Single Task}                                                                                                       \\
					\hline
					\multicolumn{1}{l}{}                  & \multicolumn{4}{c}{Warm up Section}                                                   & \multicolumn{4}{c}{Song Practice Section}                                                  \\
					\hline
					\multicolumn{1}{l}{}                  & Kernels                     & Accuracy (\%) & K value                & Accuracy(\%) & Kernels                     & Accuracy(\%) & K value                & Accuracy(\%) \\
					\hline
					learn vs play                                   & \multirow{4}{*}{Linar}      & 52.62    & \multirow{4}{*}{K = 1} & 54       & \multirow{4}{*}{Linar}      & 53.79    & \multirow{4}{*}{K = 1} & 52.41    \\
					learn vs feedback                                   &                             & 53.38    &                        & 50.13    &                             & 53.1     &                        & 51.72    \\
					play vs feedback                                   &                             & 47.5     &                        & 50.38    &                             & 54.31    &                        & 50.86    \\
					learn vs play vs feedback                                 &                             & 35.08    &                        & 36.25    &                             & 35.52    &                        & 36.55    \\
					\hline
					learn vs play                                   & \multirow{4}{*}{Polynomial} & 49       & \multirow{4}{*}{K = 3} & 50.25    & \multirow{4}{*}{Polynomial} & 53.79    & \multirow{4}{*}{K = 3} & 50.69    \\
					learn vs feedback                                 &                             & 50.75    &                        & 50.13    &                             & 50.86    &                        & 50.34    \\
					play vs feedback                                   &                             & 49.87    &                        & 49.5     &                             & 49.14    &                        & 52.07    \\
					learn vs play vs feedback                                 &                             & 33.92    &                        & 35.83    &                             & 34.71    &                        & 35.29    \\
					\hline
					learn vs play                                   & \multirow{4}{*}{RBF}        & 54.38    & \multirow{4}{*}{K = 5} & 48.37    & \multirow{4}{*}{RBF}        & 50.86    & \multirow{4}{*}{K = 5} & 50.17    \\
					learn vs feedback                                   &                             & 55.75    &                        & 52.75    &                             & 53.97    &                        & 50.17    \\
					play vs feedback                                   &                             & 51.12    &                        & 50       &                             & 53.79    &                        & 52.93    \\
					learn vs play vs feedback                                 &                             & 36.83    &                        & 34.17    &                             & 34.83    &                        & 33.1    \\
					\hline
				\end{tabular}
				
			}
			\label{tab2}
		\end{center}
	\end{table*}
	\begin{table*}[]
		
		\begin{center}
			\caption{Classification rate in children learn, children play and robot feedback across
				warm up (S1) and music practice (S2) sessions.}
			\vspace{3mm}
			\label{tab3}
			\begin{tabular}{lcccccc}
				\multicolumn{1}{c}{\multirow{2}{*}{}} & \multicolumn{3}{c}{Accuracy of SVM} & \multicolumn{3}{c}{Accuracy of KNN} \\
				\hline
				\multicolumn{1}{c}{}                  & Linear   & Polynomial    & RBF    & K = 1   & K = 3     & K = 5     \\
				\hline
				learn 1 vs learn 2                                 & 73.45    & 69.31   & 80.86  & 73.28   & 71.03   & 65      \\
				\hline
				play 1 vs play 2                                 & 75.34    & 68.79   & 80     & 74.48   & 69.14   & 64.31   \\
				\hline
				feedback 1 vs feedback 2                                 & 76.38    & 69.48   & 80.34  & 74.14   & 69.14   & 66.9   
			\end{tabular}
			
			\label{tab3}
		\end{center}
	\end{table*}
	
	\begin{table}[]
		\label{tab4}
		
		\begin{center}
			\caption{TD vs ASD Emotion Changes from Baseline and Exit Sessions.}
			\vspace{3mm}
			\begin{tabular}{llllll}
				& \textbf{Linear} & \textbf{Polynomial} & \textbf{RBF} \\
				\hline
				
				\textbf{Accuracy}                          & 75              & 62.5                & 80           \\
				\hline
				\multirow{2}{*}{\textbf{Confusion Matrix}} & 63  37          & 50  50              & 81  19       \\
				& 12  88          & 25  75              & 25  75       \\
				\hline
			\end{tabular}
			\label{tab4}
		\end{center}
	\end{table}
	
	In the first part of the analysis, EDA signals were segmented into small event-based pieces according to the number of "conversations" in each section. One "conversation" was defined by three movements: 1) robot/participant demonstrates the note(s) to play; 2) robot/participant repeats the note(s); and 3) robot/participant presents the result. Each segmentation lasts about 45 seconds. The CWT of the data, assuming the use of the C-Morlet wavelet function, was used inside a frequency range of (0.5, 50)Hz. A Support Vector Machines (SVM) classifier was then employed to classify "conversation" segmentation among three sections using the wavelet-based features. Table \ref{tab1}shows the classification accuracy for the SVM classifier with different kernel functions. As seen in this table, emotion arousal change between warmup (S1) and music practice (S2) and S2 and music game (S3) sections can be classified using a wavelet-based feature extraction SVM classifier with an average accuracy of 76\% and 70\%, respectively. With the highest percentage of accuracy for S1 and S3 being 64\%, fewer emotion changes between the S1 and S3 sections may be indicated. 
	
	In the second part of the analysis, EDA signals were segmented into small event-based pieces according to the number of "conversations" in each section as mentioned previously. In order to discover the emotion fluctuation inside one task, each "conversation" section was carefully divided into 3 segments, as described before. Again, each segmentation lasts about 45 seconds.
	 Each segment lasts about 10 - 20 seconds. Table \ref{tab2} shows the full result of emotion fluctuation in the warm-up (S1) and music practice (S2) sections from the intervention session. Notice that all of the segments cannot be appropriately classified using the existing method. Both SVM and KNN show stable results. This may suggest that the ASD group has less emotion fluctuation or arousal change once the task starts, despite varying activities in it. Stable emotion arousal in a single task could also benefit from the proper activity content, including robot agents playing music and language used during the conversation. Friendly voice feedback was based on the performance delivered to participants who were well prepared and stored in memory. Favorable feedback occurred while receiving correct input and in the case of incorrect playing, participants were given encouragement. Since emotion fluctuation can affect learning progress, less arousal change indicates the design of intervention sessions that are robust. 
	
	Cross-section comparison is also presented below. Since each "conversation" contains 3 segments, it is necessary to have specific segments from one task to compare with the other task it corresponded to. Table \ref{tab3} shows the classification rate in children learn, children play, and robot feedback across warm-up (S1) and music practice (S2) sessions. By using RBF kernel, wavelet-based SVM classification rate had an accuracy of ~80\% for all 3 comparisons. This result also matches the result from Table \ref{tab1}. 
	
	The types of activities and procedures between the baseline and exit session for both groups were the same. Using the "conversation" concept above, each were segmented. Comparing with target and control groups using the same classifier, an accuracy rate of 80\% for detecting different groups was found (see Table \ref{tab3}). Video annotators also reported "unclear" in reading facial expressions from the ASD group. Taken together, these results suggest that even with the same activities, TD and ASD groups display different bio-reactions. It has also been reported that a significant improvement of music performance was shown in the ASD group (see Table \ref{tab4}), although both groups had similar performance at their baseline sessions. Furthermore, the TD group was found to be more motivated to improve and ultimately perfect their performance, even when they made mistakes.
	
	When comparing emotion patterns from baseline and exit sessions between TD and ASD groups in Table \ref{tab3}, differences can be found. This may suggest that we have discovered a potential way of using biosignals to help diagnose autism at an early age. According to annotators and observers, TD participants showed a strong passion for this research. Excitement, stress, and disappointment were easy to recognize and label when watching the recorded videos. On the other hand, limited facial expression changes were detected in the ASD group. This makes it challenging to determine whether the ASD participants had different feelings or had the same feelings but different biosignal activities compared to the TD group.
	
	\section{Discussion, Conclusion and Future Work}
	\ As shown by others \cite{lagasse2019assessing,lim2011effects} as well as this study, playing music to children with ASD in therapy sessions has a positive impact on improving their social communication skills. Compared to the state of the art \cite{dvir2020body,bhat2013review, corbett2008brief}, where they utilized both recorded and live music in interventional sessions for single and multiple participants, our proposed robot-based music-therapy platform is a promising intervention tool in improving social behaviors, such as motor control and turn-taking skills. As our human robot interaction studies (over 200 sessions were conducted with children), most of the participants were able to complete motor control tasks with ~70\% accuracy and 6 out of 9 participants demonstrated stable turn-taking behavior when playing music. The emotion classification SVM classifier presented illustrated that emotion arousal from the ASD group can be detected and well recognized via EDA signals. 
	
	The automated music detection system created a self-adjusting environment for participants in early sessions. Most of the ASD participants began to develop the strike movement in the initial two intervention sessions; some even mastered the motor ability throughout the very first warm-up event. The robot was able to provide verbal directions and demonstrations to participants by providing voice command input as applicable, however, the majority of the participants did not request this feedback, and instead just focused on playing with NAO. This finding suggests that the young ASD population can learn fine motor control ability from specific, well-designed activities.
	
	The purpose of using a music therapy scenario as the main activity in the current research was to create an opportunity to practice a natural turn-taking behavior during social interaction. Observing all experimental sessions, six out of nine participants exhibited proper turn-taking after one or two sessions, suggesting the practice helped improve this behavior. Specifically, participant 107 significantly improved in the last few sessions when comparing results of the baseline and final session. Participant 109 had trouble focusing on listening to the robot most of the time, however, with prompting from the researcher, they performed better at the music turn-taking activity for a short period of time. For practicing turn-taking skills, fun, motivating activities, such as the described music therapy sessions incorporating individual song preference, should be designed for children with autism.
	
	During the latter half of the sessions, participants started to recognize their favorite songs. Even though the difficulty for playing proper notes was much higher, over half of the participants became more focused in the activities. Upon observations, it became clear that older participants spent more time interacting with the activities during the song practice session compared to younger participants, especially during the half/whole song play sessions. This could be for several reasons. First, the more complex the music, the more challenging it is, and the more concentration participants need to be successful. Older individuals may be more willing to accept the challenge and are better able to enjoy the sense of accomplishment they receive from their verbal feedback at the end of each session. Prior music knowledge could also be another reason for this result as older participants may have had more opportunities to learn music at school. 
	
	The game section of each session reflected the highest interest level, not only because it was relaxing and fun to play, but also because it was an opportunity for the participants to challenge the robot to mirror their free play. This exciting phenomenon could be seen as a game of "revenge." Participant 106 exhibited this behavior by spending a significant amount of time in free play game mode. According to the session executioner and video annotators, this participant, the only girl, showed very high level of involvement for all the activities, including free play. Based on the conversation and music performance with the robot, participant 106 showed a strong interest in challenging the robot in a friendly way. High levels of engagement further supports the idea that the proposed robot-based music-therapy platform is a viable interventional tool. Additionally, these findings also highlight the need for fun and motivating activities, such as music games, to be incorporated into interventions designed for children with autism.
	
	Conducting emotion studies with children with autism can be difficult and bio-signals provide a possible way to work around the unique challenges of this population. The event-based emotion classification method adopted in this research suggests that not only are physiological signals a good tool to detect emotion in ASD populations, but it is also possible to recognize and categorize emotions using this technique. The same activity with different intensities can cause emotion change in the arousal dimension, although, for the ASD group, it is difficult to label emotions based on facial expression changes in the video annotation phase. Fewer emotion fluctuations in a particular activity, as seen in Table \ref{tab2}, suggests that a mild, friendly, game-like therapy system may encourage better social content learning for children with autism, even when there are repetitive movements. These well-designed activities could provide a relaxed learning environment that helps participants focus on learning music content with proper communication behaviors. This may explain the improvement in music play performance during the song practice (S2) section of intervention sessions, as seen in Figure \ref{fig10}.
	
	There were several limitations of this study that should be considered. As is the case with most research studies, a larger sample size is needed to better understand the impact of this therapy on children with ASD. The ASD group in this paper also only included one female. Future work should include more females to better portray and understand an already underrepresented portion of this population of interest. In addition, improving skills as complex as turn-taking for example, likely requires more sessions over a longer period of time to better enhance the treatment and resulting behavioral changes. Music practice can also be tedious if it is the only activity in an existing interaction system and may have influenced the degree of interaction in which participants engaged with the robot. Future research could include more activities or multiple instruments, as opposed to just one, to further diversify sessions. The choice of instrument, and its resulting limitations, may also have an impact. The Xylophone, for example, is somewhat static and future research could modify or incorporate a different instrument that can produce more melodies, accommodate more complex songs, and portray a wider range of musical emotion and expression. Finally, another limitation of this study was the type of communication the robot and participant engaged in. Verbal communication was not rich between participants and NAO, and instead, most of the time, participants could follow the instructions from the robot without asking it for help. While this was likely not an issue for the specific non-verbal behavioral goals of this intervention, future work should expand upon this study by incorporating other elements of behavior that often have unique deficits in this population, including speech.
	
	In summary, this paper presented a novel robot-based music-therapy platform to model and improve social behaviors in children with ASD. In addition to the novel platform introduced, this study also incorporated emotion recognition and classification utilizing EDA physiological signals of arousal. Results of this study are consistent with findings in the literature for TD and ASD children and suggest that the proposed platform is a viable tool to facilitate the improvement of fine motor and turn-taking skills in children with ASD.
	
	\section*{Acknowledgment}
	\ This research was partially supported by gifts for research on autism to the University of Denver from several family members. The researchers would like to thank the children and their family members for their dedication and willingness to volunteer their time and make this research possible.
	
	\bibliographystyle{frontiersinSCNS_ENG_HUMS} 
	\bibliography{ref}

\begin{thebibliography}{49}
\providecommand{\natexlab}[1]{#1}
\expandafter\ifx\csname urlstyle\endcsname\relax
  \providecommand{\doi}[1]{doi:\discretionary{}{}{}#1}\else
  \providecommand{\doi}{doi:\discretionary{}{}{}\begingroup
  \urlstyle{rm}\Url}\fi
\providecommand{\selectlanguage}[1]{\relax}
\providecommand{\bibAnnoteFile}[1]{%
  \IfFileExists{#1}{\begin{quotation}\noindent\textsc{Key:} #1\\
  \textsc{Annotation:}\ \input{#1}\end{quotation}}{}}
\providecommand{\bibAnnote}[2]{%
  \begin{quotation}\noindent\textsc{Key:} #1\\
  \textsc{Annotation:}\ #2\end{quotation}}

\bibitem[{Anzalone et~al.(2015)Anzalone, Boucenna, Ivaldi, and
  Chetouani}]{anzalone2015evaluating}
Anzalone, S.~M., Boucenna, S., Ivaldi, S., and Chetouani, M. (2015).
\newblock Evaluating the engagement with social robots.
\newblock \emph{International Journal of Social Robotics} 7, 465--478
\bibAnnoteFile{anzalone2015evaluating}

\bibitem[{Anzalone et~al.(2014)Anzalone, Tilmont, Boucenna, Xavier, Jouen,
  Bodeau et~al.}]{anzalone2014children}
Anzalone, S.~M., Tilmont, E., Boucenna, S., Xavier, J., Jouen, A.-L., Bodeau,
  N., et~al. (2014).
\newblock How children with autism spectrum disorder behave and explore the
  4-dimensional (spatial 3d+ time) environment during a joint attention
  induction task with a robot.
\newblock \emph{Research in Autism Spectrum Disorders} 8, 814--826
\bibAnnoteFile{anzalone2014children}

\bibitem[{Askari et~al.(2018)Askari, Feng, Sweeny, and
  Mahoor}]{askari2018pilot}
Askari, F., Feng, H., Sweeny, T.~D., and Mahoor, M.~H. (2018).
\newblock A pilot study on facial expression recognition ability of autistic
  children using ryan, a rear-projected humanoid robot.
\newblock In \emph{2018 27th IEEE International Symposium on Robot and Human
  Interactive Communication (RO-MAN)} (IEEE), 790--795
\bibAnnoteFile{askari2018pilot}

\bibitem[{Association(2000)}]{DSMIV2000}
Association, A.~P. (2000).
\newblock Diagnostic and statistical manual of mental disorders: Dsm-iv
\bibAnnoteFile{DSMIV2000}

\bibitem[{Beer et~al.(2016)Beer, Boren, and Liles}]{beer2016robot}
Beer, J.~M., Boren, M., and Liles, K.~R. (2016).
\newblock Robot assisted music therapy a case study with children diagnosed
  with autism.
\newblock In \emph{2016 11th ACM/IEEE International Conference on Human-Robot
  Interaction (HRI)} (IEEE), 419--420
\bibAnnoteFile{beer2016robot}

\bibitem[{Bhat and Srinivasan(2013)}]{bhat2013review}
Bhat, A.~N. and Srinivasan, S. (2013).
\newblock A review of “music and movement” therapies for children with
  autism: embodied interventions for multisystem development.
\newblock \emph{Frontiers in integrative neuroscience} 7, 22
\bibAnnoteFile{bhat2013review}

\bibitem[{Boccanfuso et~al.(2016)Boccanfuso, Barney, Foster, Ahn, Chawarska,
  Scassellati et~al.}]{boccanfuso2016emotional}
Boccanfuso, L., Barney, E., Foster, C., Ahn, Y.~A., Chawarska, K., Scassellati,
  B., et~al. (2016).
\newblock Emotional robot to examine different play patterns and affective
  responses of children with and without asd.
\newblock In \emph{2016 11th ACM/IEEE International Conference on Human-Robot
  Interaction (HRI)} (IEEE), 19--26
\bibAnnoteFile{boccanfuso2016emotional}

\bibitem[{Boso et~al.(2007)Boso, Emanuele, Minazzi, Abbamonte, and
  Politi}]{boso2007effect}
Boso, M., Emanuele, E., Minazzi, V., Abbamonte, M., and Politi, P. (2007).
\newblock Effect of long-term interactive music therapy on behavior profile and
  musical skills in young adults with severe autism.
\newblock \emph{The journal of alternative and complementary medicine} 13,
  709--712
\bibAnnoteFile{boso2007effect}

\bibitem[{Boucenna et~al.(2014{\natexlab{a}})Boucenna, Anzalone, Tilmont,
  Cohen, and Chetouani}]{boucenna2014learning}
Boucenna, S., Anzalone, S., Tilmont, E., Cohen, D., and Chetouani, M.
  (2014{\natexlab{a}}).
\newblock Learning of social signatures through imitation game between a robot
  and a human partner.
\newblock \emph{IEEE Transactions on Autonomous Mental Development} 6, 213--225
\bibAnnoteFile{boucenna2014learning}

\bibitem[{Boucenna et~al.(2016)Boucenna, Cohen, Meltzoff, Gaussier, and
  Chetouani}]{boucenna2016robots}
Boucenna, S., Cohen, D., Meltzoff, A.~N., Gaussier, P., and Chetouani, M.
  (2016).
\newblock Robots learn to recognize individuals from imitative encounters with
  people and avatars.
\newblock \emph{Scientific reports} 6, 1--10
\bibAnnoteFile{boucenna2016robots}

\bibitem[{Boucenna et~al.(2014{\natexlab{b}})Boucenna, Narzisi, Tilmont,
  Muratori, Pioggia, Cohen et~al.}]{boucenna2014interactive}
Boucenna, S., Narzisi, A., Tilmont, E., Muratori, F., Pioggia, G., Cohen, D.,
  et~al. (2014{\natexlab{b}}).
\newblock Interactive technologies for autistic children: A review.
\newblock \emph{Cognitive Computation} 6, 722--740
\bibAnnoteFile{boucenna2014interactive}

\bibitem[{Brownell(2002)}]{brownell2002musically}
Brownell, M.~D. (2002).
\newblock Musically adapted social stories to modify behaviors in students with
  autism: Four case studies.
\newblock \emph{Journal of music therapy} 39, 117--144
\bibAnnoteFile{brownell2002musically}

\bibitem[{Cibrian et~al.(2020)Cibrian, Madrigal, Avelais, and
  Tentori}]{cibrian2020supporting}
Cibrian, F.~L., Madrigal, M., Avelais, M., and Tentori, M. (2020).
\newblock Supporting coordination of children with asd using neurological music
  therapy: A pilot randomized control trial comparing an elastic touch-display
  with tambourines.
\newblock \emph{Research in Developmental Disabilities} 106, 103741
\bibAnnoteFile{cibrian2020supporting}

\bibitem[{Constantino and Gruber(2012)}]{constantino2012social}
Constantino, J.~N. and Gruber, C.~P. (2012).
\newblock \emph{Social responsiveness scale: SRS-2} (Western psychological
  services Torrance, CA)
\bibAnnoteFile{constantino2012social}

\bibitem[{Corbett et~al.(2008)Corbett, Shickman, and Ferrer}]{corbett2008brief}
Corbett, B.~A., Shickman, K., and Ferrer, E. (2008).
\newblock Brief report: the effects of tomatis sound therapy on language in
  children with autism.
\newblock \emph{Journal of autism and developmental disorders} 38, 562--566
\bibAnnoteFile{corbett2008brief}

\bibitem[{Dellatan(2003)}]{dellatan2003use}
Dellatan, A.~K. (2003).
\newblock The use of music with chronic food refusal: A case study.
\newblock \emph{Music Therapy Perspectives} 21, 105--109
\bibAnnoteFile{dellatan2003use}

\bibitem[{Di~Nuovo et~al.(2018)Di~Nuovo, Conti, Trubia, Buono, and
  Di~Nuovo}]{di2018deep}
Di~Nuovo, A., Conti, D., Trubia, G., Buono, S., and Di~Nuovo, S. (2018).
\newblock Deep learning systems for estimating visual attention in
  robot-assisted therapy of children with autism and intellectual disability.
\newblock \emph{Robotics} 7, 25
\bibAnnoteFile{di2018deep}

\bibitem[{Diehl et~al.(2012)Diehl, Schmitt, Villano, and
  Crowell}]{diehl2012clinical}
Diehl, J.~J., Schmitt, L.~M., Villano, M., and Crowell, C.~R. (2012).
\newblock The clinical use of robots for individuals with autism spectrum
  disorders: A critical review.
\newblock \emph{Research in autism spectrum disorders} 6, 249--262
\bibAnnoteFile{diehl2012clinical}

\bibitem[{Dvir et~al.(2020)Dvir, Lotan, Viderman, and Elefant}]{dvir2020body}
Dvir, T., Lotan, N., Viderman, R., and Elefant, C. (2020).
\newblock The body communicates: Movement synchrony during music therapy with
  children diagnosed with asd.
\newblock \emph{The Arts in Psychotherapy} 69, 101658
\bibAnnoteFile{dvir2020body}

\bibitem[{Feng et~al.(2018)Feng, Golshan, and Mahoor}]{feng2018wavelet}
Feng, H., Golshan, H.~M., and Mahoor, M.~H. (2018).
\newblock A wavelet-based approach to emotion classification using eda signals.
\newblock \emph{Expert Systems with Applications} 112, 77--86
\bibAnnoteFile{feng2018wavelet}

\bibitem[{Feng et~al.(2013)Feng, Gutierrez, Zhang, and Mahoor}]{feng2013can}
Feng, H., Gutierrez, A., Zhang, J., and Mahoor, M.~H. (2013).
\newblock Can nao robot improve eye-gaze attention of children with high
  functioning autism?
\newblock In \emph{2013 IEEE International Conference on Healthcare
  Informatics} (IEEE), 484--484
\bibAnnoteFile{feng2013can}

\bibitem[{Gifford et~al.(2011)Gifford, Srinivasan, Kaur, Dotov, Wanamaker,
  Dressler et~al.}]{gifford2011using}
Gifford, T., Srinivasan, S., Kaur, M., Dotov, D., Wanamaker, C., Dressler, G.,
  et~al. (2011).
\newblock Using robots to teach musical rhythms to typically developing
  children and children with autism.
\newblock \emph{University of Connecticut}
\bibAnnoteFile{gifford2011using}

\bibitem[{Guedjou et~al.(2017)Guedjou, Boucenna, Xavier, Cohen, and
  Chetouani}]{guedjou2017influence}
Guedjou, H., Boucenna, S., Xavier, J., Cohen, D., and Chetouani, M. (2017).
\newblock The influence of individual social traits on robot learning in a
  human-robot interaction.
\newblock In \emph{2017 26th IEEE International Symposium on Robot and Human
  Interactive Communication (RO-MAN)} (IEEE), 256--262
\bibAnnoteFile{guedjou2017influence}

\bibitem[{Han et~al.(2008)Han, Ma, and Huang}]{han2008novel}
Han, X.-A., Ma, Y., and Huang, X. (2008).
\newblock A novel generalization of b{\'e}zier curve and surface.
\newblock \emph{Journal of Computational and Applied Mathematics} 217, 180--193
\bibAnnoteFile{han2008novel}

\bibitem[{Kappas et~al.(2013)Kappas, K{\"u}ster, Basedow, and
  Dente}]{kappas2013validation}
Kappas, A., K{\"u}ster, D., Basedow, C., and Dente, P. (2013).
\newblock A validation study of the affectiva q-sensor in different social
  laboratory situations.
\newblock In \emph{53rd Annual Meeting of the Society for Psychophysiological
  Research, Florence, Italy}. vol.~39
\bibAnnoteFile{kappas2013validation}

\bibitem[{Kim et~al.(2013)Kim, Berkovits, Bernier, Leyzberg, Shic, Paul
  et~al.}]{kim2013social}
Kim, E.~S., Berkovits, L.~D., Bernier, E.~P., Leyzberg, D., Shic, F., Paul, R.,
  et~al. (2013).
\newblock Social robots as embedded reinforcers of social behavior in children
  with autism.
\newblock \emph{Journal of autism and developmental disorders} 43, 1038--1049
\bibAnnoteFile{kim2013social}

\bibitem[{Kim et~al.(2008)Kim, Wigram, and Gold}]{kim2008effects}
Kim, J., Wigram, T., and Gold, C. (2008).
\newblock The effects of improvisational music therapy on joint attention
  behaviors in autistic children: a randomized controlled study.
\newblock \emph{Journal of autism and developmental disorders} 38, 1758
\bibAnnoteFile{kim2008effects}

\bibitem[{LaGasse et~al.(2019)LaGasse, Manning, Crasta, Gavin, and
  Davies}]{lagasse2019assessing}
LaGasse, A.~B., Manning, R.~C., Crasta, J.~E., Gavin, W.~J., and Davies, P.~L.
  (2019).
\newblock Assessing the impact of music therapy on sensory gating and attention
  in children with autism: a pilot and feasibility study.
\newblock \emph{Journal of music therapy} 56, 287--314
\bibAnnoteFile{lagasse2019assessing}

\bibitem[{Lim and Draper(2011)}]{lim2011effects}
Lim, H.~A. and Draper, E. (2011).
\newblock The effects of music therapy incorporated with applied behavior
  analysis verbal behavior approach for children with autism spectrum
  disorders.
\newblock \emph{Journal of music therapy} 48, 532--550
\bibAnnoteFile{lim2011effects}

\bibitem[{Loomes et~al.(2017)Loomes, Hull, and Mandy}]{loomes2017male}
Loomes, R., Hull, L., and Mandy, W. P.~L. (2017).
\newblock What is the male-to-female ratio in autism spectrum disorder? a
  systematic review and meta-analysis.
\newblock \emph{Journal of the American Academy of Child \& Adolescent
  Psychiatry} 56, 466--474
\bibAnnoteFile{loomes2017male}

\bibitem[{Marinoiu et~al.(2018)Marinoiu, Zanfir, Olaru, and
  Sminchisescu}]{marinoiu20183d}
Marinoiu, E., Zanfir, M., Olaru, V., and Sminchisescu, C. (2018).
\newblock 3d human sensing, action and emotion recognition in robot assisted
  therapy of children with autism.
\newblock In \emph{Proceedings of the IEEE conference on computer vision and
  pattern recognition}. 2158--2167
\bibAnnoteFile{marinoiu20183d}

\bibitem[{Mavadati et~al.(2014)Mavadati, Feng, Gutierrez, and
  Mahoor}]{mavadati2014comparing}
Mavadati, S.~M., Feng, H., Gutierrez, A., and Mahoor, M.~H. (2014).
\newblock Comparing the gaze responses of children with autism and typically
  developed individuals in human-robot interaction.
\newblock In \emph{2014 IEEE-RAS International Conference on Humanoid Robots}
  (IEEE), 1128--1133
\bibAnnoteFile{mavadati2014comparing}

\bibitem[{Mihalache et~al.(2020)Mihalache, Feng, Askari, Sokol-Hessner, Moody,
  Mahoor et~al.}]{mihalache2020perceiving}
Mihalache, D., Feng, H., Askari, F., Sokol-Hessner, P., Moody, E.~J., Mahoor,
  M.~H., et~al. (2020).
\newblock Perceiving gaze from head and eye rotations: An integrative challenge
  for children and adults.
\newblock \emph{Developmental Science} 23, e12886
\bibAnnoteFile{mihalache2020perceiving}

\bibitem[{Molnar-Szakacs and Heaton(2012)}]{molnar2012music}
Molnar-Szakacs, I. and Heaton, P. (2012).
\newblock Music: a unique window into the world of autism.
\newblock \emph{Annals of the New York Academy of Sciences} 1252, 318--324
\bibAnnoteFile{molnar2012music}

\bibitem[{M{\"o}ssler et~al.(2019)M{\"o}ssler, Gold, A{\ss}mus, Schumacher,
  Calvet, Reimer et~al.}]{mossler2019therapeutic}
M{\"o}ssler, K., Gold, C., A{\ss}mus, J., Schumacher, K., Calvet, C., Reimer,
  S., et~al. (2019).
\newblock The therapeutic relationship as predictor of change in music therapy
  with young children with autism spectrum disorder.
\newblock \emph{Journal of autism and developmental disorders} 49, 2795--2809
\bibAnnoteFile{mossler2019therapeutic}

\bibitem[{Pennisi et~al.(2016)Pennisi, Tonacci, Tartarisco, Billeci, Ruta,
  Gangemi et~al.}]{pennisi2016autism}
Pennisi, P., Tonacci, A., Tartarisco, G., Billeci, L., Ruta, L., Gangemi, S.,
  et~al. (2016).
\newblock Autism and social robotics: A systematic review.
\newblock \emph{Autism Research} 9, 165--183
\bibAnnoteFile{pennisi2016autism}

\bibitem[{Reschke-Hern{\'a}ndez(2011)}]{reschke2011history}
Reschke-Hern{\'a}ndez, A.~E. (2011).
\newblock History of music therapy treatment interventions for children with
  autism.
\newblock \emph{Journal of Music Therapy} 48, 169--207
\bibAnnoteFile{reschke2011history}

\bibitem[{Richardson et~al.(2018)Richardson, Coeckelbergh, Wakunuma, Billing,
  Ziemke, Gomez et~al.}]{richardson2018robot}
Richardson, K., Coeckelbergh, M., Wakunuma, K., Billing, E., Ziemke, T., Gomez,
  P., et~al. (2018).
\newblock Robot enhanced therapy for children with autism (dream): A social
  model of autism.
\newblock \emph{IEEE Technology and Society Magazine} 37, 30--39
\bibAnnoteFile{richardson2018robot}

\bibitem[{Roper(2003)}]{roper2003melodic}
Roper, N. (2003).
\newblock Melodic intonation therapy with young children with apraxia.
\newblock \emph{Bridges} 1, 1--7
\bibAnnoteFile{roper2003melodic}

\bibitem[{Scassellati et~al.(2012)Scassellati, Admoni, and
  Matari{\'c}}]{scassellati2012robots}
Scassellati, B., Admoni, H., and Matari{\'c}, M. (2012).
\newblock Robots for use in autism research.
\newblock \emph{Annual review of biomedical engineering} 14
\bibAnnoteFile{scassellati2012robots}

\bibitem[{Starr and Zenker(1998)}]{starr1998understanding}
Starr, E. and Zenker, K. (1998).
\newblock Understanding autism in the context of music therapy: Bridging theory
  and practice.
\newblock \emph{Canadian Journal of Music Therapy}
\bibAnnoteFile{starr1998understanding}

\bibitem[{Stephens(2008)}]{stephens2008spontaneous}
Stephens, C.~E. (2008).
\newblock Spontaneous imitation by children with autism during a repetitive
  musical play routine.
\newblock \emph{Autism} 12, 645--671
\bibAnnoteFile{stephens2008spontaneous}

\bibitem[{Taheri et~al.(2015)Taheri, Alemi, Meghdari, Pouretemad, Basiri, and
  Poorgoldooz}]{taheri2015impact}
Taheri, A., Alemi, M., Meghdari, A., Pouretemad, H., Basiri, N.~M., and
  Poorgoldooz, P. (2015).
\newblock Impact of humanoid social robots on treatment of a pair of iranian
  autistic twins.
\newblock In \emph{International Conference on Social Robotics} (Springer),
  623--632
\bibAnnoteFile{taheri2015impact}

\bibitem[{Taheri et~al.(2019)Taheri, Meghdari, Alemi, and
  Pouretemad}]{taheri2019teaching}
Taheri, A., Meghdari, A., Alemi, M., and Pouretemad, H. (2019).
\newblock Teaching music to children with autism: a social robotics challenge.
\newblock \emph{Scientia Iranica} 26, 40--58
\bibAnnoteFile{taheri2019teaching}

\bibitem[{Taheri et~al.(2016)Taheri, Meghdari, Alemi, Pouretemad, Poorgoldooz,
  and Roohbakhsh}]{taheri2016social}
Taheri, A., Meghdari, A., Alemi, M., Pouretemad, H., Poorgoldooz, P., and
  Roohbakhsh, M. (2016).
\newblock Social robots and teaching music to autistic children: Myth or
  reality?
\newblock In \emph{International Conference on Social Robotics} (Springer),
  541--550
\bibAnnoteFile{taheri2016social}

\bibitem[{Thaut and Clair(2000)}]{thaut2000scientific}
Thaut, M.~H. and Clair, A.~A. (2000).
\newblock \emph{A scientific model of music in therapy and medicine} (IMR
  Press)
\bibAnnoteFile{thaut2000scientific}

\bibitem[{VandenBos(2007)}]{vandenbos2007american}
VandenBos, G. (2007).
\newblock American psychological association dictionary.
\newblock \emph{Washington, DC: American Psychological Association}
\bibAnnoteFile{vandenbos2007american}

\bibitem[{Warwick and Alvin(1991)}]{warwick1991music}
Warwick, A. and Alvin, J. (1991).
\newblock \emph{Music therapy for the autistic child} (Oxford University Press)
\bibAnnoteFile{warwick1991music}

\bibitem[{Zheng et~al.(2015)Zheng, Young, Swanson, Weitlauf, Warren, and
  Sarkar}]{zheng2015robot}
Zheng, Z., Young, E.~M., Swanson, A.~R., Weitlauf, A.~S., Warren, Z.~E., and
  Sarkar, N. (2015).
\newblock Robot-mediated imitation skill training for children with autism.
\newblock \emph{IEEE Transactions on Neural Systems and Rehabilitation
  Engineering} 24, 682--691
\bibAnnoteFile{zheng2015robot}

\end{thebibliography}
	
\end{document}